\documentclass[%
 reprint,
 amsmath,amssymb,
 aps,
 twocolumn,
 superscriptaddress,
]{revtex4-1}

\usepackage{amssymb, latexsym, amsopn}

\usepackage{graphicx}
\usepackage{dcolumn}
\usepackage{bm}
\usepackage{mathrsfs}
\usepackage{enumitem}
\usepackage[mathlines]{lineno}

\usepackage{outlines}

\usepackage{tabularx}

\usepackage{xcolor}

\usepackage{float}
\usepackage[caption = false]{subfig}

\begin{document}

\preprint{APS/123-QED}

\title{All-particle cosmic ray energy spectrum measured\\
by the HAWC experiment from 10 to 500 TeV}

\author{R.~Alfaro}\affiliation{Instituto de F{\'i}sica, Universidad Nacional Aut{\'o}noma de M{\'e}xico, Mexico City, Mexico}
\author{C.~Alvarez}\affiliation{Universidad Aut{\'o}noma de Chiapas, Tuxtla Guti{\'e}rrez, Chiapas, Mexico}
\author{R.~Arceo}\affiliation{Universidad Aut{\'o}noma de Chiapas, Tuxtla Guti{\'e}rrez, Chiapas, Mexico}
\author{J.C.~Arteaga-Vel{\'a}zquez}\affiliation{Universidad Michoacana de San Nicol{\'a}s de Hidalgo, Morelia, Mexico}
\author{D.~Avila Rojas}\affiliation{Instituto de F{\'i}sica, Universidad Nacional Aut{\'o}noma de M{\'e}xico, Mexico City, Mexico}
\author{H.A.~Ayala Solares}\affiliation{Department of Physics, Michigan Technological University, Houghton, MI, USA}
\author{A.S.~Barber}\affiliation{Department of Physics and Astronomy, University of Utah, Salt Lake City, UT, USA}
\author{A.~Becerril}\affiliation{Instituto de F{\'i}sica, Universidad Nacional Aut{\'o}noma de M{\'e}xico, Mexico City, Mexico}
\author{E.~Belmont-Moreno}\affiliation{Instituto de F{\'i}sica, Universidad Nacional Aut{\'o}noma de M{\'e}xico, Mexico City, Mexico}
\author{S.Y.~BenZvi}\affiliation{Department of Physics \& Astronomy, University of Rochester, Rochester, NY, USA}
% ...

\collaboration{HAWC Collaboration}\noaffiliation


\begin{abstract}
We report on the measurement of the all-particle cosmic ray
energy spectrum with the High Altitude Water Cherenkov
(HAWC) Observatory in the energy range 10 to 500 TeV.
HAWC is a ground based air-shower array deployed
on the slopes of Volcan Sierra Negra in the state of Puebla, Mexico,
and is sensitive to gamma rays and cosmic rays at TeV energies.
The data used in this work were taken from 234 days between June 2016 to February 2017.
The primary cosmic-ray energy is determined with a maximum likelihood approach using 
the particle density as a function of distance to the shower core. 
Introducing quality cuts to isolate events with shower cores landing on the array, 
the reconstructed energy distribution is unfolded iteratively.
The measured all-particle spectrum is consistent with a broken power law with an index
of $-2.49\pm0.01$ prior to a break at $(45.7\pm0.1$) TeV, followed by an index of $-2.71\pm0.01$.
The spectrum also respresents a single measurement that spans the energy range 
between direct detection and ground based experiments.
As a verification of the detector response, the energy scale and angular resolution 
are validated by observation of the cosmic ray Moon shadow's dependence on energy.
\end{abstract}

\pacs{Valid PACS appear here}
\maketitle

\section{Introduction} \label{intro}
The primary cosmic ray spectrum spans a wide range covering over ten decades in 
energy and thirty decades in flux \cite{horandel:gcr, engel:gcr}.
It is well described by a nearly single power law of index $\gamma \approx -2.7$,
with two prominent breaks at $\sim3\times10^{15}$ eV and $\sim3\times10^{18}$ eV, 
referred to as the \textit{knee} and \textit{ankle}, respectively.
The non-thermal nature of the spectrum, including the breaks, carries information
regarding the dynamics of the environments in which cosmic rays are accelerated and those that they traverse.
Despite significant experimental efforts to study this structure, 
our understanding of the nature of acceleration sites remains incomplete.

For cosmic ray particles with energy below about 100 TeV, direct measurements from satellite and balloon-borne
experiments provide the most detailed measurements of the primary particle spectrum.
Recent results from the PAMELA \cite{pamela} satellite demonstrate a decrease in the 
proton to helium flux ratio up to TV rigidities or $\sim$ TeV energies. 
Similarly, the CREAM balloon-borne detector \cite{cream:2009, cream:2010, cream:2011} reports further hardening 
of the helium energy spectrum, surpassing the proton flux at approximately 10 TeV.
The flattening of the helium flux relative to protons also has been reported by the ATIC experiment \cite{atic-2:2009}.
This type of structure could be an indication of different source populations \cite{zatsepin:three_comp_model},
or a nearby source that is proton rich up to TeV energies \cite{tomassetti:2015}.
It has also been suggested that the spectral hardening could be attributed to anomalous diffusion \cite{pamela:analysis_p_he}.

Probing the cosmic ray spectrum via direct detection becomes a challenge in the 10--100 TeV 
range and beyond due to limited detector exposures fueled by a rapidly diminishing flux.
Ground-based air shower arrays are not as limited by their collection area but are not as sensitive
to the identity of the primary particle,
as well as being reliant on simulations to estimate extensive air shower parameters.
Yet, as demonstrated by recent results from the ARGO-YBJ \cite{argo:all-particle}, GRAPES-3 \cite{grapes-iii}
and Tibet-III \cite{tibet:all-particle} experiments, ground-based arrays are best suited to probe
the all-particle flux above 100 TeV. 

The energy response to hadronic air showers of the HAWC Observatory \cite{hawc:crab}
allows for a detailed measurement of the cosmic ray flux above 10 TeV.
This is due in part to the similar size of multi-TeV showers and the containment area of HAWC, 
as well as the array's proximity to shower maximum at these energies, where shower fluctuations are minimized.
This places the HAWC observatory in a position to bridge cosmic ray measurements 
between direct detection apparatuses and larger PeV-scale air-shower array experiments.
In this paper, we demonstrate that ability by reporting on the 
analysis of the data sample collected by the HAWC
experiment in the period from June 2016 -- February 2017 and the measurement of
the all-particle cosmic ray energy spectrum between 10--500 TeV.
The cosmic ray Moon shadow's evolution with energy is presented
as a verification of the energy scale in these measurements.

The paper is organized as follows: in section \ref{hawcobs} 
the main characteristics of the HAWC detector are described.
The simulation methods and event sample used for the analysis are described in section \ref{mcsample}.
Section \ref{analysis} is devoted to the data analysis including details of the event selection,
event reconstruction, the unfolding method, and sources of systematic uncertainties.
The results are presented in section \ref{results} and discussed in section \ref{discussion}.
Section \ref{conclusion} summarizes the main conclusions of this work.

\begin{center}
 \begin{figure*}[!htb]
  \centering
    \subfloat{
              \includegraphics[width=0.5\linewidth,trim={0 0 -4cm 0}]{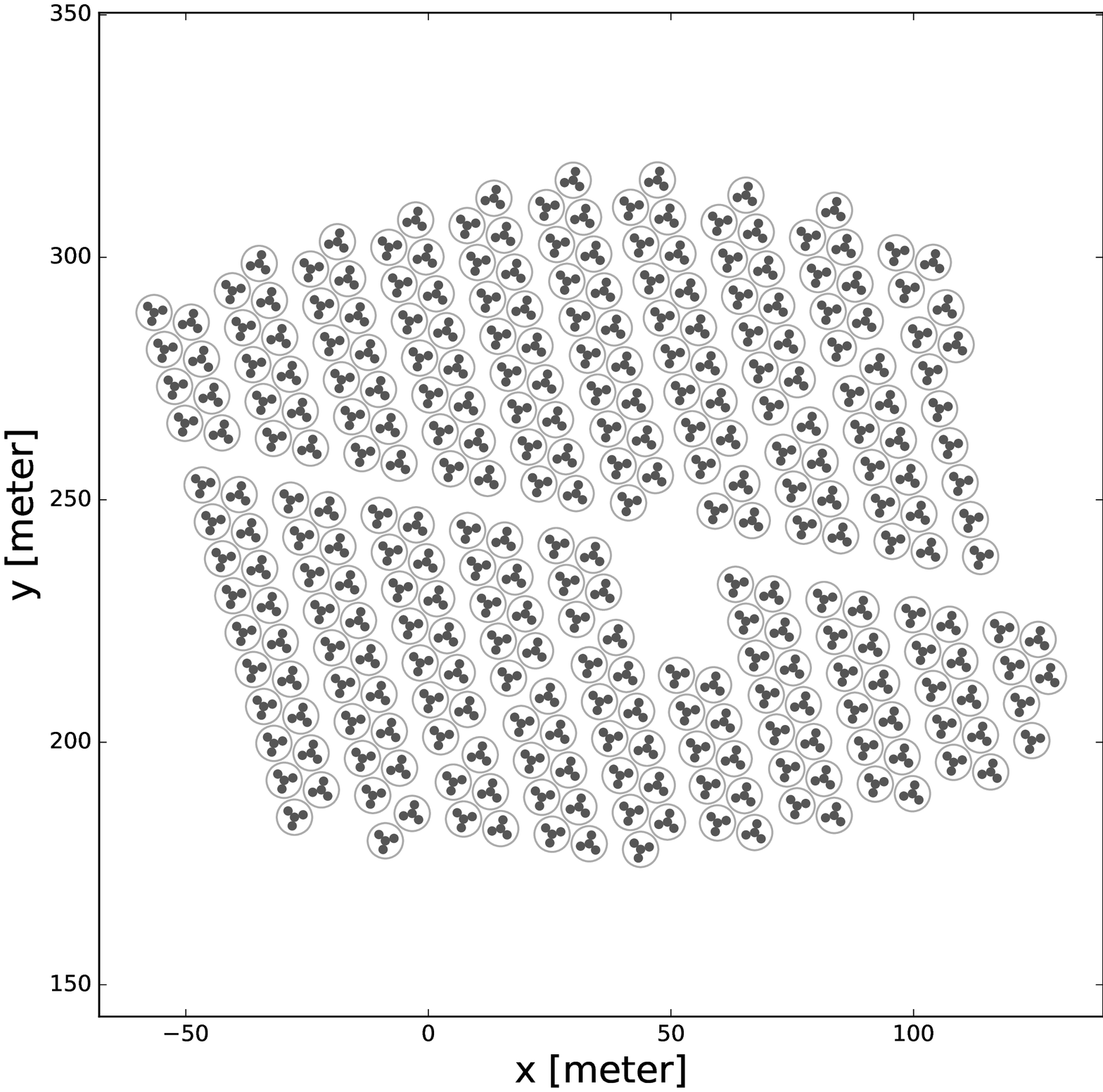}
             }
    \subfloat{
              \includegraphics[width=0.3\linewidth,trim={0 -2cm 0 0},clip]{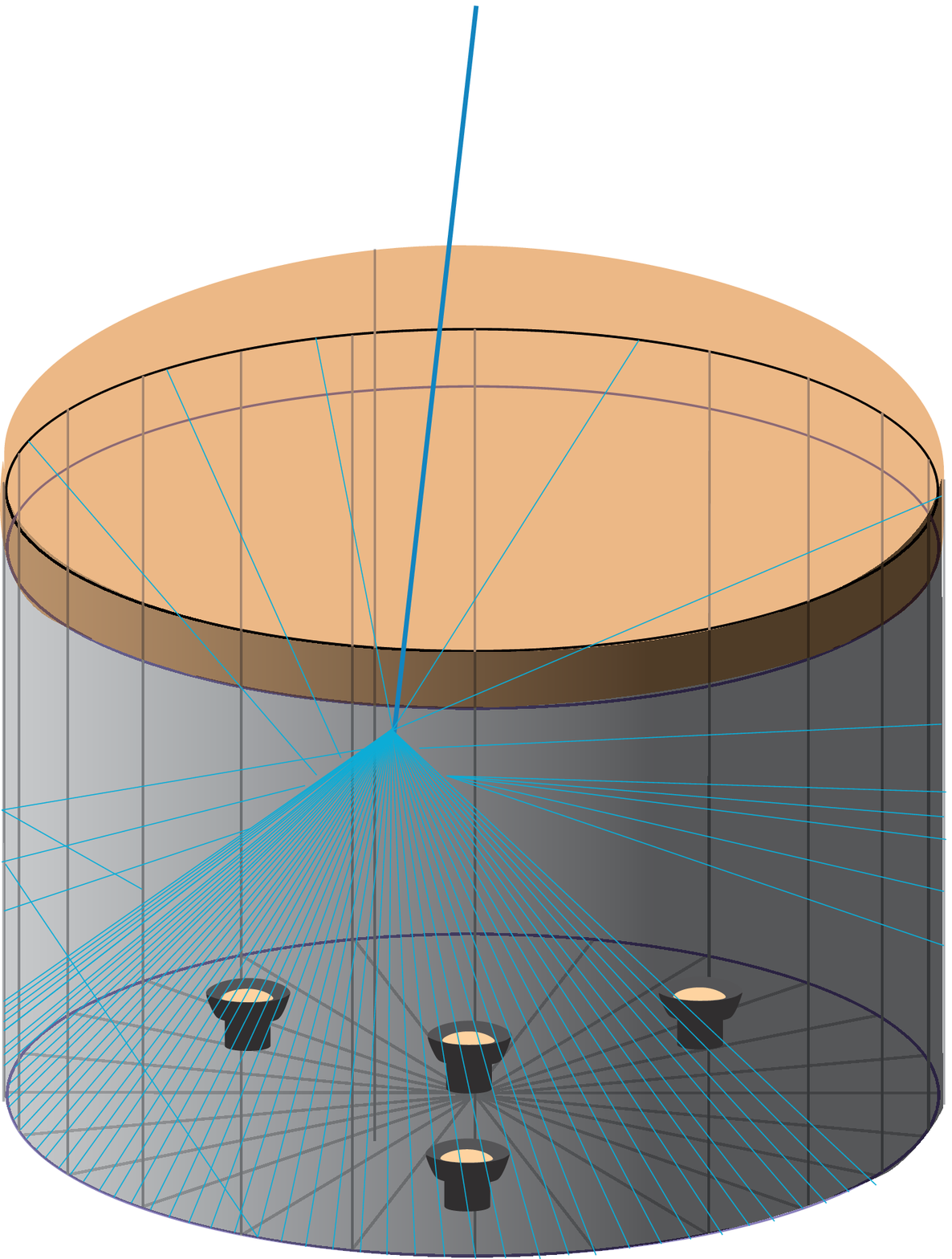}
             }
  
    \caption{The left panel shows the layout of the entire HAWC detector.
    Each WCD is indicated by a large circle encompassing the smaller, darker circles which identify the PMTs.
    The right panel depicts the representation of a single WCD including the steel tank, the protective roof, and the four PMTs.
    The penetrating dark blue line represents a high-energy secondary air shower particle, 
    which emits Cherenkov radiation indicated by the cyan rays inside the WCD volume.}
    \label{fig:hawc_diag}
  \end{figure*}
\end{center}

\section{The HAWC Observatory and Data Selection} \label{hawcobs}
The Earth's atmosphere is opaque to cosmic rays in the energy range of HAWC. 
Primary cosmic ray particles interact with air molecules to produce large cascades of 
secondary particles, called extensive air showers.
Properties of the primary cosmic ray must be inferred from the 
air shower particles that reach ground level.
The HAWC Observatory is an air shower array located at 4100 m a.s.l.
on the slopes of Volcan Sierra Negra in the state of Puebla, Mexico.
HAWC is designed to detect air showers produced by primary 
gamma rays in the 500 GeV to 100 TeV energy range, 
but its altitude and physical dimensions permit 
measurements of primary hadronic cosmic rays up to multi-PeV energies.

The detector comprises a 22,000 m$^2$ array of 294 close-packed Water 
Cherenkov Detectors (WCDs).
Each WCD consists of a 4.5 m tall and 7.3 m diameter cylindrical steel tank 
lined with a black plastic bladder and filled with 188,000 liters of purified water. 
Attached to the floor of each tank are four upward-facing photo-multiplier tubes (PMTs): 
one central high-quantum efficiency Hamamatsu 10-inch R7081 PMT, 
and three 8-inch R5912 PMTs each at 1.8 m from the center forming an equilateral triangle.
The PMTs observe the Cherenkov light produced when secondary particles 
(primarily electrons, positrons and gamma rays) from air showers enter the tank.
Figure \ref{fig:hawc_diag} depicts the full HAWC array and the schematic of a WCD.

The PMT signals are transferred via RG59 coaxial cables 
to a central counting house, where they are amplified, shaped 
and discriminated on custom front-end boards using two voltage thresholds: 
one at $1/4$ and the other at $4$ photoelectrons (PEs).
The time stamps when these thresholds are crossed are recorded and
provide a means of inferring the amplitude of the measured signal.
The resulting Time-over-Threshold (ToT) is proportional 
to the logarithm of the pulse's total charge.
Commercial Time-to-Digital Converters (TDCs) digitize the ToTs,
and send the data to a farm of computers for further processing.
A simple multiplicity trigger is used to identify candidate 
air shower events, ensuring that a minimum number of
PMTs record signals within a defined time window.

The event reconstruction procedure involves determination of air shower properties 
including local arrival direction, core position, and an estimate of the primary energy.
While a cursory reconstruction is performed on-site, 
this analysis uses the results from the fourth revision (Pass-4) of the off-site reconstruction,
to have a uniform data set and the most updated calibrations available.
The calibration procedure permits the estimation of the true number of PEs in a PMT from the measured ToT.
It is performed by an on-site laser system that sends pulses to each WCD while the PMT responses are recorded.
A further calibration step is required to account for varying cable lengths resulting in PMT timing differences,
which are determined to sub-ns precision.

The HAWC detector in its full configuration was in stable data taking mode during the 
runs selected for this analysis, amounting to a total live time of 234 days from 8 June, 2016 to 17 February, 2017.
The total up-time efficiency was $\sim 92\%$ and the mean trigger rate was $\sim25$ kHz.
All events triggering at least 75 PMTs and passing the core and angle fitting routines were processed for final analysis.
Further event selection described in section \ref{eventselection} identifies a data set consisting of 
$8.42 \times 10^{9}$ events with a mean energy of $\sim3$ TeV.

\section{Simulation} \label{mcsample}
\subsection{Air Shower Events}

\subsubsection{Extensive Air Showers} \label{eassim}

Extensive air showers are characterized by a laterally extended 
but thin disk of secondary particles.
The nature of the primary particle determines the evolution of the shower
with regard to the particle content and subsequently the shape and 
energy distribution as the shower develops through the atmosphere.

For a primary particle interacting in the atmosphere, the resulting 
air shower is simulated using the CORSIKA \cite{Heck:1998vt} package (v740), 
with FLUKA \cite{Battistoni:2007, Ferrari:2005}, and QGSJet-II-03 \cite{qgsjet}
as the low energy and high energy particle physics interaction models, respectively.
Smaller simulation sets were generated for hadronic interaction systematic studies 
using the EPOS (LHC) \cite{epos} and SIBYLL 2.1 \cite{sibyll} high energy models.

Primaries of the eight species measured by the CREAM flights \cite{cream:2009,cream:2011} 
(H, $^{4}$He, $^{12}$C, $^{16}$O, $^{20}$Ne, $^{24}$Mg, $^{28}$Si, $^{56}$Fe)
were generated on an $E^{-2}$ differential energy spectrum from 5 GeV -- 3 PeV 
and distributed over a 1 km radius circular throwing area.
The simulated zenith angle range was $0^{\circ}$--$70^{\circ}$, azimuthally symmetric, 
and weighted to a $\sin{\theta}\cos{\theta}$ arrival distribution.
The secondary charged particles interacting with the HAWC detector were simulated using the GEANT4 \cite{Agostinelli:2002hh} package.
Dedicated HAWC software was used to simulate the detector response over the entire hardware and data analysis chain.

\subsubsection{Composition} \label{compsim}

Comparison of the measured Moon shadow with predictions as well as the unfolding of the 
all-particle spectrum requires an assumption on the composition of the cosmic-ray spectrum.
The spectra of individual primary elements were weighted according to a broken power law of the form:
\begin{equation} \label{eq:fitFunc}
 \begin{split}
  \mathscr{F}(A, E_{\text{br}}, & \gamma_1, \gamma_2) =\\&
    \begin{cases}
      A \cdot \left( \frac{E}{E_{0}} \right) ^{\gamma_1}  & \ \ E < E_{\text{br}} \\
      A \cdot \left( \frac{E_{\text{br}}}{E_{0}} \right)^{\gamma_1-\gamma_2} \cdot \left( \frac{E}{E_{0}} \right)^{\gamma_2} & \ \ E \ge E_{\text{br}} \, ,
    \end{cases}
   \end{split}
\end{equation}
where $A$ is the normalization at $E_{0}$ with spectral index $\gamma_1$, and 
a second index $\gamma_2$ starting at energy $E_{\text{br}}$.
For all species's fits, $E_{0}=100$ GeV.
The nominal composition assumption used in simulation is the best fit of equation
\ref{eq:fitFunc} to direct measurement data 
provided by AMS \cite{ams:proton,ams:helium}, CREAM \cite{cream:2009,cream:2011}, 
and PAMELA \cite{pamela}.
The fluxes for proton and helium were allowed to vary independently while
those for elements with atomic number $Z > 2$ were varied together and thus 
have the same spectral indices.
The resulting fit values for the various species are presented in table \ref{tab:crfits_table},
defining the nominal composition model used in this analysis.
\renewcommand{\arraystretch}{1.6}
\begin{center}
 \begin{table}[!htb]
   \caption{
            Equation \ref{eq:fitFunc} parameters for the hadronic species considered for the assumed composition in this analysis.
            The parameters were obtained as best fits to AMS \cite{ams:proton,ams:helium}, CREAM \cite{cream:2009,cream:2011}, 
            and PAMELA \cite{pamela} data.
            Uncertainties in the fits were included in estimating the systematic uncertainties of flux measurements
            due to composition assumptions.
           }
   \begin{tabularx}{0.48\textwidth}{ c  c  c  c  c }
     \hline
     \hline
     & $A$ & $E_{\text{br}}$ & $-\gamma_1$ & $-\gamma_2$ \\ 
     & [GeV s sr m$^{2}$]$^{-1}$ & [GeV] & & \\ 
     \hline
     H  & $(4.48\pm0.04)\times10^{-2}$ & $440.6^{+87.8}_{-62.7}$ & $2.81\pm0.01$ & $2.66\pm0.01$\\
     He & $(3.31\pm0.02)\times10^{-2}$ & $854.8^{+125.7}_{-105.5}$ & $2.73\pm0.01$ & $2.54\pm0.01$ \\
     C  & $(6.96\pm0.18)\times10^{-6}$ & $2882^{+904.4}_{-481.9}$ & $2.76\pm0.03$ & $2.55\pm0.04$ \\
     O  & $(5.00\pm0.09)\times10^{-6}$ & $3843^{+1206}_{-643}$    & $2.76\pm0.03$ & $2.55\pm0.04$ \\
     Ne & $(6.31\pm0.35)\times10^{-7}$ & $4803^{+1507}_{-803}$   & $2.76\pm0.03$ & $2.55\pm0.04$ \\
     Mg & $(5.70\pm0.26)\times10^{-7}$ & $5764^{+1809}_{-964}$   & $2.76\pm0.03$ & $2.55\pm0.04$ \\
     Si & $(5.70\pm0.13)\times10^{-7}$ & $6725^{+2110}_{-1124}$  & $2.76\pm0.03$ & $2.55\pm0.04$ \\
     Fe & $(2.00\pm0.04)\times10^{-7}$ & $13450^{+4220}_{-2249}$ & $2.76\pm0.03$ & $2.55\pm0.04$ \\
     \hline
     \hline
   \end{tabularx}
   \label{tab:crfits_table}
 \end{table}
\end{center}

\begin{center}
 \begin{figure*}[!htb]
  \centering
    \subfloat{
              \includegraphics[height=0.3\textwidth,trim={0 0.25cm 2cm 2cm},clip]{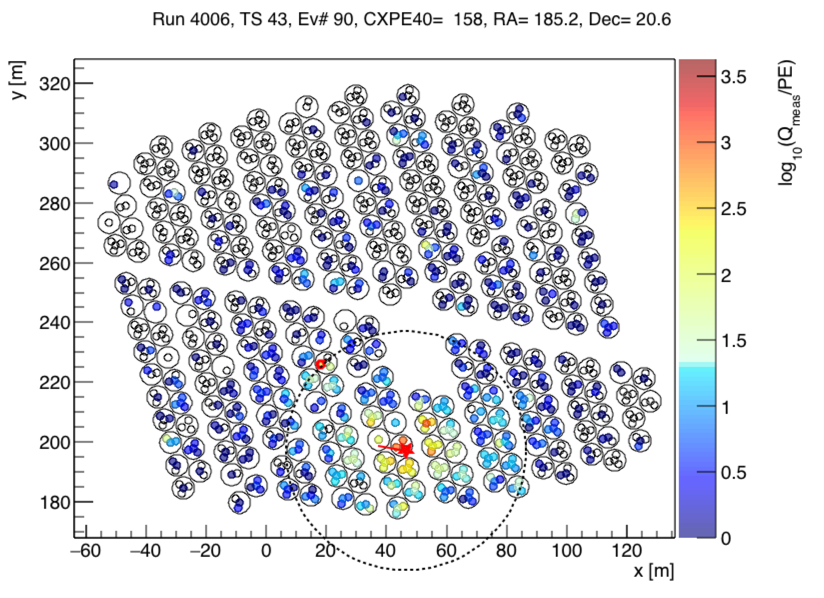}
             }
    \subfloat{
              \includegraphics[height=0.3\textwidth,trim={1cm 0.25cm 2cm 2cm},clip]{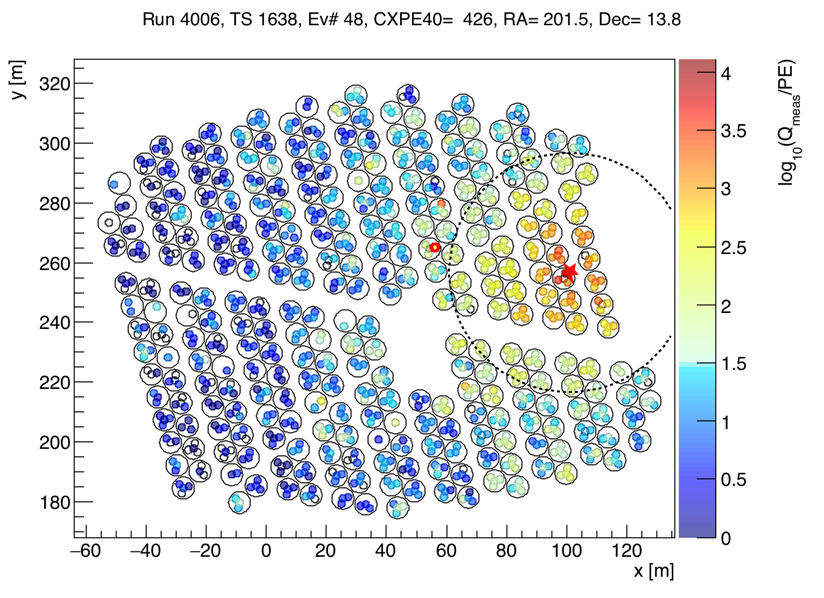}
             }
  
    \caption[Sample air shower events]{
             Example air shower events from the data sample, where the color scale indicates
             the logarithm of the measured signal in PEs.
             The core location is indicated by the red star, and the 40 meter 
             circle shows the region used for $N_{\text{r40}}$.
             The reconstructed energy of the events are 10 TeV (left) and 100 TeV (right),
             where the uncertainty on each is $0.1$ in $\log{E_{\text{reco}}}$.
             }
    \label{fig:samp-events}
  \end{figure*}
\end{center}

\subsection{Moon Shadow} \label{moonsim}
The Moon blocks the nearly isotropic flux of cosmic rays on their path to Earth, 
and because of their charge, cosmic rays interact with the Earth's magnetic field
which bends their trajectories.
The deflection is dependent on both the particle charge and energy, so this 
has the effect of shifting the shadow in relation to the Moon's true position.
Using an accurate representation of the geomagnetic field, we can compare simulated 
trajectories to the observed cosmic ray Moon shadow as an independent test of the detector's 
angular resolution and energy response.

To simulate the cosmic ray Moon shadow, we developed a charged particle propagation routine
including geomagnetic effects using the International Geomagnetic Reference Field (IGRF) \cite{igrf2015} model.
To efficiently generate sufficient statistics to model the shadow, we used the set of 345 graphics processing units (GPUs)
available at the Wisconsin IceCube Particle Astrophysics Center.
The OpenCL API specification \cite{opencl} is used for the GPU kernel invocation, 
with the host central processing unit (CPU) backend being written with PyOpenCL \cite{pyopencl}.
Each GPU is capable of simulating the propagation of $\sim10^6$ particles nearly simultaneously, 
providing statistics $O(10^9)$ within hour time scales, which translates to a speedup of $90\times$
over available CPU resources.

Particles are backtraced from the location of the HAWC detector until reaching the orbital radius of the Moon,
and intersection with the Moon sphere is determined.
The Boris-Buneman method \cite{boris,plasmacomp} is used to integrate the relativistic equations of motion, 
due to its symplecticity and wide use in plasma physics applications. 
Furthermore, we employ a fixed angular displacement step of $10^{-3}$ $\text{rad}$ by sampling the field strength 
at each numerical integration, serving as an adaptive step in time.
This ensures that in regions of higher magnetic field strength where charged particle deflection is greater, 
the particle trajectory is sampled more finely than where the field strength is weaker.
If the maximum allowable number of integration steps ($10^4$) is reached,
the particle is assumed to be below the geomagnetic cutoff and not considered further.

\section{Analysis} \label{analysis}
To infer shower properties from the raw hit data, 
events are subject to a reconstruction procedure.
The quantities of interest are the shower \textit{direction} and \textit{core location}, 
which are then used to estimate the primary energy, $E_{\textnormal{reco}}$.
A set of selection criteria is applied to improve the quality of the reconstructed 
event sample and minimize potential bias in our estimate of the detector response.
We validate the simulated detector response by the observation of the cosmic ray 
Moon shadow's evolution with energy.
The all-particle energy spectrum is determined using an iterative unfolding procedure,
and sources of systematic uncertainties are taken into account.

\subsection{Direction and Core Reconstruction}
The front of particles in extensive air showers assumes a conical form whose 
main axis defines the arrival direction and whose apex is the shower core.
The highest density of secondary particles coincides with the core location.
Farther away, the shower front becomes less dense and wider in time.
For the electromagnetic component of an extensive air shower, the expected particle density as 
a function of the lateral distance from the core is well approximated by the 
Nishimura-Kamata-Greisen (NKG) function \cite{nkg}.
We implement a modification that includes a Gaussian component, 
such that the signal amplitude $S$ given a core position $\vec{x}$ at a point $\vec{x}_i$ is given by
\begin{gather}
 \begin{split}
   S(A, \vec{x}, \vec{x}_i) &=\\ 
              & A \cdot \Bigg(\frac{1}{2\pi \sigma^2}e^{\frac{-|\vec{x}_i - \vec{x}|^2}{2\sigma^2}} 
              + \frac{N}{(0.5 + \frac{|\vec{x}_i-\vec{x}|}{R_{m}})^3}\Bigg) \, ,
 \end{split}
 \label{eq:sfcf}
\end{gather}
where $A$ is a normalization factor, 
$R_{m}$ is the Moli\`ere radius of the atmosphere (approximately 120 m at HAWC 
altitude), $\sigma$ is the width of the Gaussian, and $N$ is the normalization of the NKG tail.
The values of $\sigma$ and $N$ are fixed to $10$ m and $5 \times 10^{-5}$, respectively.
The modification ameliorates observed excessive iterations due to the NKG function's 
rapidly increasing amplitude near $r=0$, as well as computationally expensive calls 
to fitting the exponential involving the age parameter.

The estimated core position $\vec{x}$ and local zenith and 
azimuth angles ($\theta,\phi$) are reconstructed in an iterative procedure described in \cite{hawc:crab}.
Using all triggered PMTs in an event, a simple center-of-mass core estimate serves as a seed 
to the more elaborate lateral distribution function presented in equation \ref{eq:sfcf}. 
This core position is then provided to a plane fit estimate of the shower arrival direction.
Provided these initial estimates of $\vec{x}, \theta, \phi$, 
a selection of PMTs within $\pm50$ ns of a curvature correction to the shower plane
is provided for a final pass of the core and angle fitters.
Two example events which have passed the event selection criteria are shown in figure \ref{fig:samp-events}.

\subsection{Energy Estimation} \label{lhee}
To estimate the primary cosmic ray energy, we use the lateral distribution of the measured signal 
as a function of the primary particle energy.
Using a proton-initiated air shower simulation, we build a four-dimensional probability table with bins 
in zenith angle, primary energy, PMT distance from the core in the shower plane (lateral distance), 
and measured PMT signal amplitude.
The table provides the probability for observing a given PMT amplitude Q at a lateral distance R
from the shower core in a proton shower of energy $E$ and zenith angle $\theta$.
The tables are smoothed with a multi-dimensional spline fitting routine \cite{photospline}
to ensure bin fluctuations from the simulation statistics do not influence energy estimation.
Given a shower with a reconstructed arrival direction and core position, 
each PMT contributes a likelihood value extracted from the tables, including operational PMTs 
that do not record a signal.
For each possible energy, the logarithms of the likelihood values for all PMTs are summed,
and the energy bin with the maximal likelihood value is chosen as the best energy estimate.

The proton energy table takes the following form:
\begin{outline}
  \1 Three zenith angle bins 
    \2 $\theta_{0}: 0.957 \le \cos \theta \le 1 $
    \2 $\theta_{1}: 0.817\le \cos \theta < 0.957$
    \2 $\theta_{2}: 0.5 \le \cos \theta < 0.817$
  \1 Forty-four energy bins from 70 GeV--1.4 PeV with bin width of $0.1$ in $\log E$
  \1 Seventy bins in lateral distance R from 0--350 m in bins of width $5$ m
  \1 Forty bins in charge Q from 1--$10^{6}$ photoelectrons (PE) in steps of $0.15$ in $\log\,$Q
\end{outline}
The lateral distance bin spacing was assigned half of the spacing between WCD centers, which is of order $10$ m.
Finally the charge bin width was chosen to be of the same order as the estimated PMT charge resolution ($\sim$ 30\%).

\begin{figure}[!h]
            \includegraphics[width=0.5\textwidth]{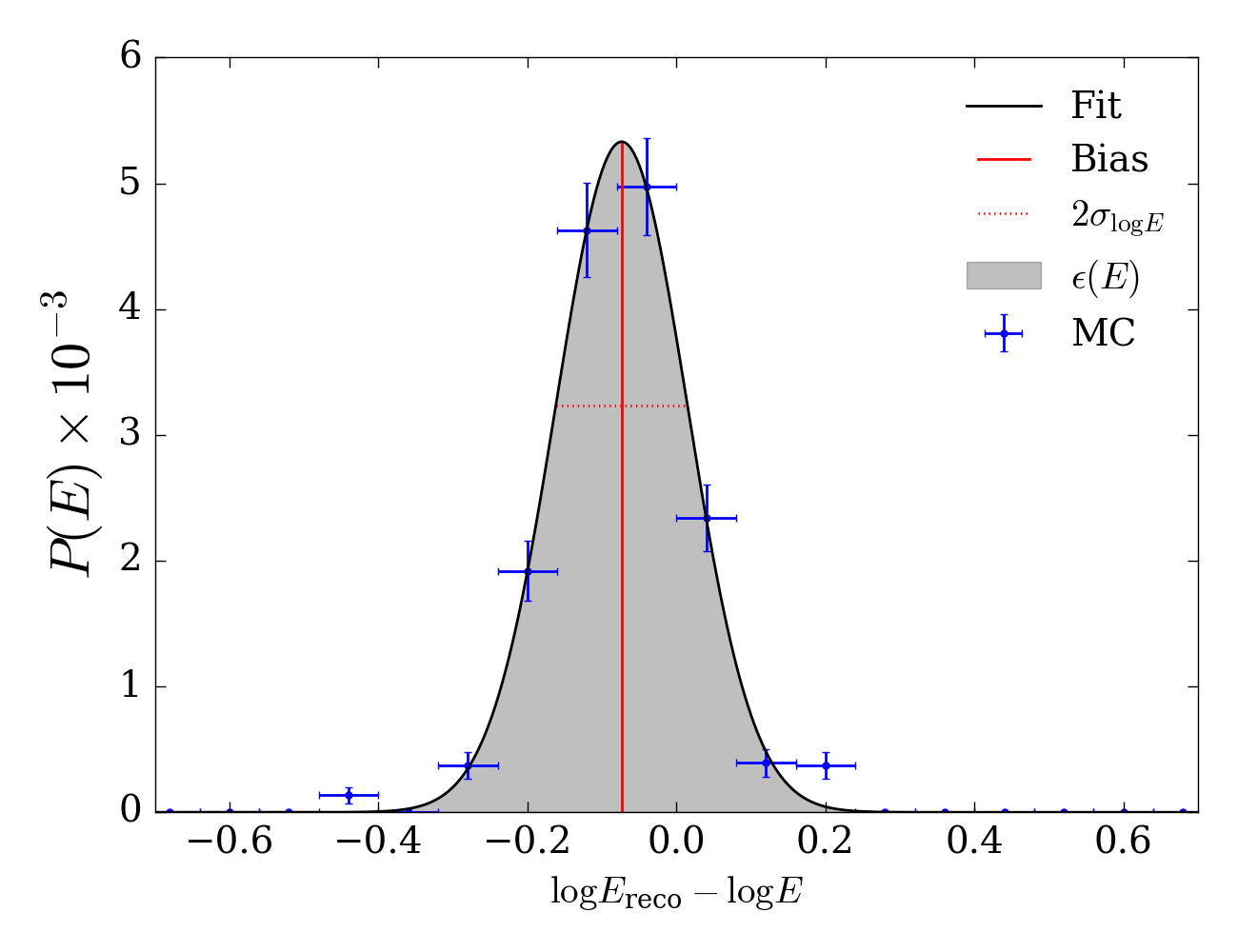}
  \caption{
           Difference between the logarithms of the reconstructed $E_{\textnormal{reco}}$ and true energy $E$ for the bin centered at
           $E=100$ TeV, showing the definitions of energy bias, resolution, and efficiency.
           The values from simulation are indicated by the blue markers,
           while the black curve is a Gaussian fit to these points.
           The normalization condition includes events that were not selected
           as a result of the selection criteria, thus, the integral represents
           the efficiency $\epsilon(E)$ to reconstruct and select events in this energy bin.
          }

  \label{fig:lhe-bias-dist}
\end{figure}

\begin{center}
 \begin{figure*}[!htb]
  \centering
   \subfloat{
             \includegraphics[width=0.5\textwidth]{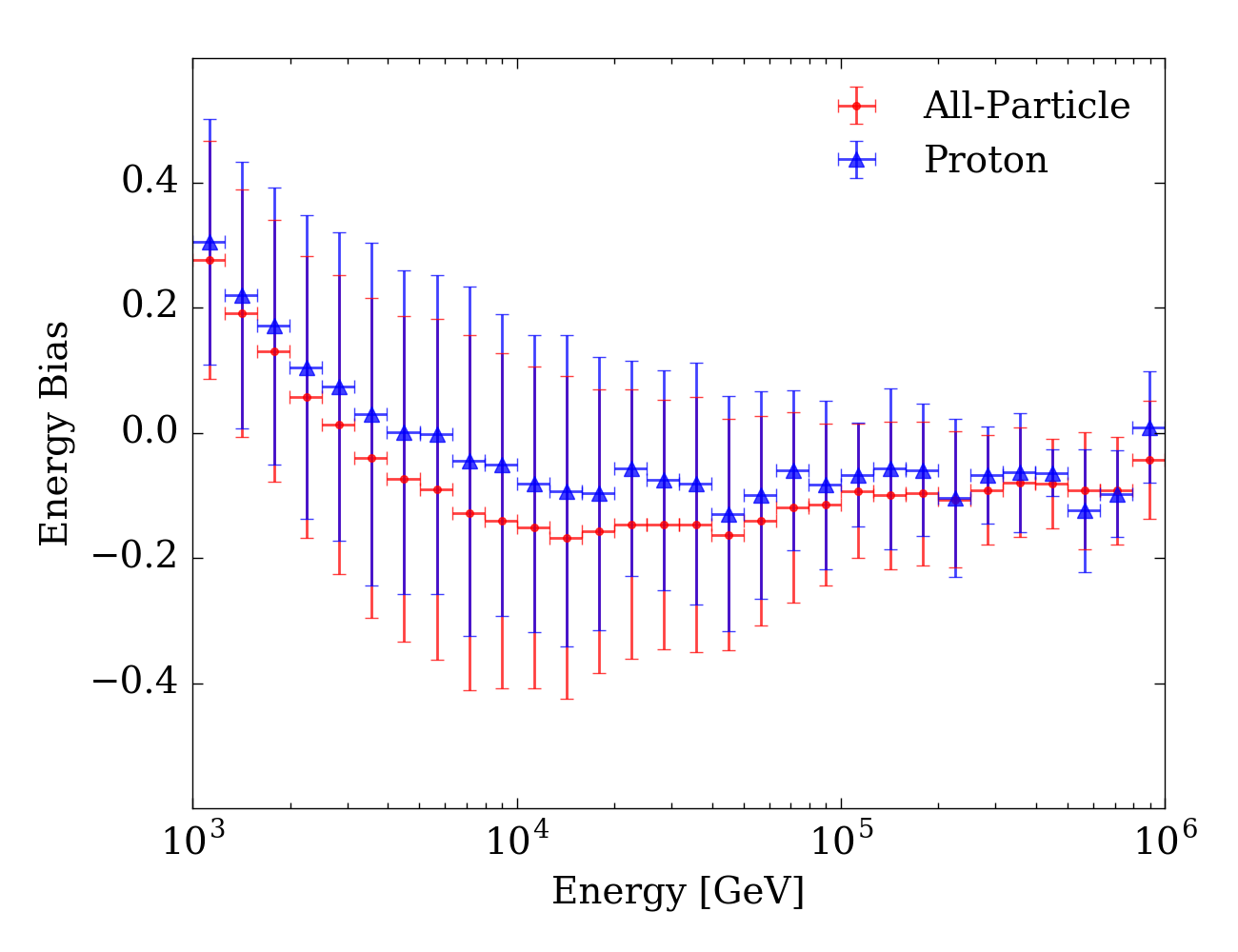}
            }
   \subfloat{
             \includegraphics[width=0.5\textwidth]{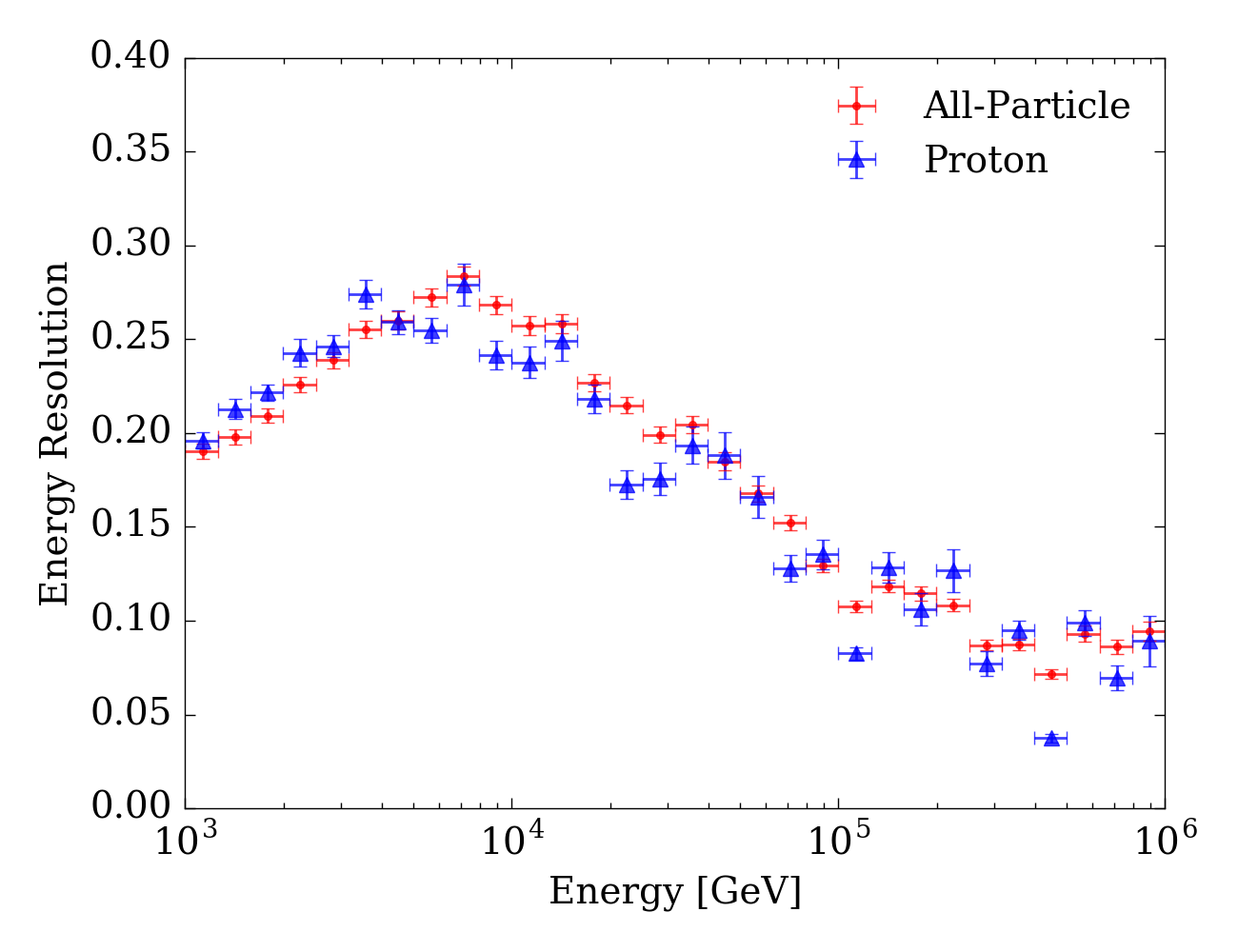}
            }
   \caption{
            Resulting energy bias (left) and resolution (right) for 
            all particles and protons as a function of the true energy using the energy 
            estimation method provided the event selection criteria.
            The vertical bars in the left panel represent the width of the bias distribution or the 
            energy resolution, while the uncertainty in the bias value is comparatively miniscule.
            The uncertainties shown for the energy resolution are those estimated 
            from the fit of a Gaussian to the bias.
            The bars in the coordinate represent the bin width in true energy.
           }
 
   \label{fig:lhe-performance}
 \end{figure*}
\end{center}

The resulting performance is evaluated via the bias distribution, defined by the 
difference between the logarithms of the reconstructed and true energy values:
\begin{equation}\label{eq:energy_bias}
    \text{bias} = \log{E_{\text{reco}}} - \log{E} \, ,
\end{equation}
shown for a single energy bin in figure \ref{fig:lhe-bias-dist}.
The mean of this distribution defines the energy bias or offset and the width defines the energy resolution,
which are shown as a function of energy in the panels of figure \ref{fig:lhe-performance},
provided the event selection criteria.
We also identify the integral of the bias distribution as the efficiency $\epsilon(E)$ to reconstruct events 
at energy $E$, as the normalization condition includes events that are not reconstructed or do not pass 
the selection criteria in that energy bin.

Given that the energy estimation tables are built solely from proton simulation, the energy bias for evaluating only proton showers 
is within half of a table bin width above $\sim4$ TeV.
This sets an energy threshold below which proton showers just passing the selection cuts will be reconstructed at higher energies,
manifesting as an increasingly larger bias with decreasing true energy.
This can also be seen as an apparent drop in the energy resolution below $\sim7$ TeV due to the decreasing sample size in evaluating
equation \ref{eq:energy_bias}.

\subsection{Event Selection} \label{eventselection}
The following selection criteria have been adopted for both simulated events and data:
\begin{enumerate}[label=(\alph*)]
  \item Events must pass the core and angle fitter, which are required for the energy reconstruction.

  \item Events must pass a minimum multiplicity threshold of $N_{\text{hit}}\ge75$ PMTs.
        This value was chosen in accordance with previous cosmic ray analyses \cite{hawc:anisotropy},
        as well as efficiency optimization for the energy reconstruction algorithm which samples the probability
        tables with $O(10^3)$ PMTs for each energy bin.

  \item Events are chosen to be within the first zenith angle bin of the energy estimation table.
        This corresponds to nearly vertical events with $\theta \le 16.7^{\circ}$,
        limiting the influence of shower development from increased atmospheric
        overburden with increasing zenith angle.

  \item At least 40 PMTs within 40 meters of the core position ($N_{\text{r40}} \ge 40$) must record a signal.
        This criterion ensures that selected events land on or within 15 meters of the array,
        resulting in an estimated core resolution of better than 10 meters above 10 TeV.
        Additionally, the resulting uncertainty in the effective area is below $10 \%$.
        More harsh values up to $N_{\text{r40}} \ge 100$ were tested to further constrain cores to the 
        array with the cost of increasing the uncertainty on the effective area;
        however the influence on the measured spectrum was less than $5 \%$.
        Values less than 20 result in core and angular resolutions above 25 m and $2^{\circ}$, respectively,
        in addition to causing energy resolutions to exceed 100\%.
\end{enumerate}
The effects of making these cuts on simulated and data events
are shown in table \ref{tab:evt_sel}.
The core resolution provided by the above selection criteria 
is estimated to be 10 m at 10 TeV, dropping below 8 m above 100 TeV.
The angular resolution above 10 TeV is better than $0.5^{\circ}$.
\renewcommand{\arraystretch}{1.3}
\begin{center}
 \begin{table}[!htb]
  \centering
  \caption{
           Passing percentages for successive application of event quality cuts in simulation and data,
           including the observed event rate in data.
           The percentages represent the fraction of events that passed the previous cut,
           with the set of triggered events being the reference selection.
          }
  \begin{tabularx}{0.48\textwidth}{c c c c}
    \hline\hline
    Cut                         & \% Passing         &                 & Data Event Rate \\
                                & MC                 & Data            & [kHz] \\
    \hline
    No cut (trig. threshold)    &  $100$ \%          & $100$  \%       & $24.7$ \\
    Core \& angle fit pass      &  $99$ \%           & $96$   \%       & $23.6$ \\
    $N_{\text{hit}}\ge 75$      &  $31$ \%           & $23$   \%       & $5.7$ \\
    $\theta < 17^{\circ}$       &  $8$ \%            & $6$   \%        & $1.5$ \\
    $N_{r40}\ge40$              &  $2$ \%            & $2$    \%       & $0.43$ \\
    \hline\hline
  \end{tabularx}
  \label{tab:evt_sel}
 \end{table}
\end{center}

\subsection{Moon Shadow} \label{moonshadow}
\begin{figure}[!h]
            \includegraphics[width=0.9\linewidth,trim={.5cm 0.5cm .5cm 1.65cm},clip]{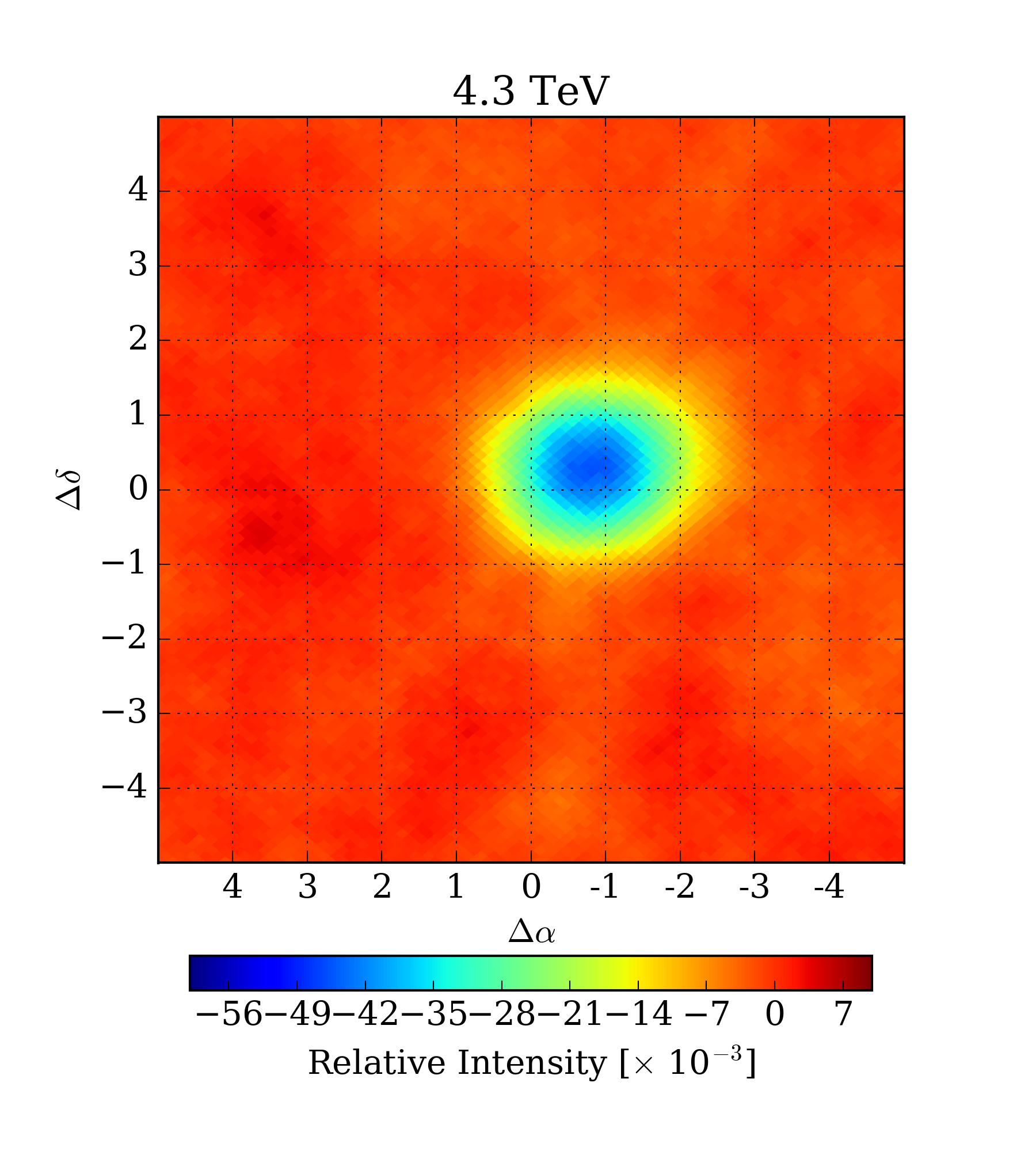}
  \caption{
           Relative intensity of the Moon shadow at a mean energy of 4.3 TeV.
           The map has been smoothed with a top-hat function by $1^{\circ}$ to enhance the shadow visually.
           A two-dimensional Gaussian was fit to the unsmoothed maps.
          }

  \label{fig:moon_shadow}
\end{figure}

We applied the event selection criteria and energy estimation technique to the 
observation of the cosmic ray Moon shadow as a test of the detector response.
The zenith angle cut was relaxed to $\theta\le45^{\circ}$ in order to obtain sufficient statistics 
for a significant observation of the shadow.
The resulting sample size is $4.2\times10^{10}$ events.
We follow the methods presented in \cite{hawc:anisotropy,hawc:dirint} for making sky maps.
Eleven Moon-centered maps were made in recontructed energy bins of width $0.2$ in $\log{E_\textnormal{reco}}$ from 1--100 TeV,
and the true energy of each bin was estimated from simulation.
The observed deficit of the Moon shadow is measured with relative intensity,
giving the amplitude of deviations from the isotropic expectation:
\begin{equation}
 \centering
  \delta I(\alpha_{i},\delta_{i}) = 
       \frac{N_{i}-\langle N \rangle_{i}}
            {\langle N \rangle_{i}} \, ,
  \label{eq:relint}
\end{equation}
where $\langle N \rangle_{i}$ is the estimated background counts and $N_{i}$ the observed
counts in bin $i$ with right ascension and declination $\alpha_{i},\delta_{i}$, respectively.
An example of the Moon shadow observed at an estimated energy of $4.3$ TeV is shown in figure \ref{fig:moon_shadow}.
For each map, the resulting Moon shadow was fit to a two-dimensional Gaussian,
from which the offset to the true Moon position in declination ($\Delta\delta$) and right ascension ($\Delta\alpha$) were evaluated.
The resulting dependency of the combined angular offset was compared to the expected
deviation from simulation, taking into account the detector response and the composition assumption defined in section \ref{compsim}.

\begin{figure}[hpt]
  \subfloat{
            \includegraphics[clip,width=\columnwidth]{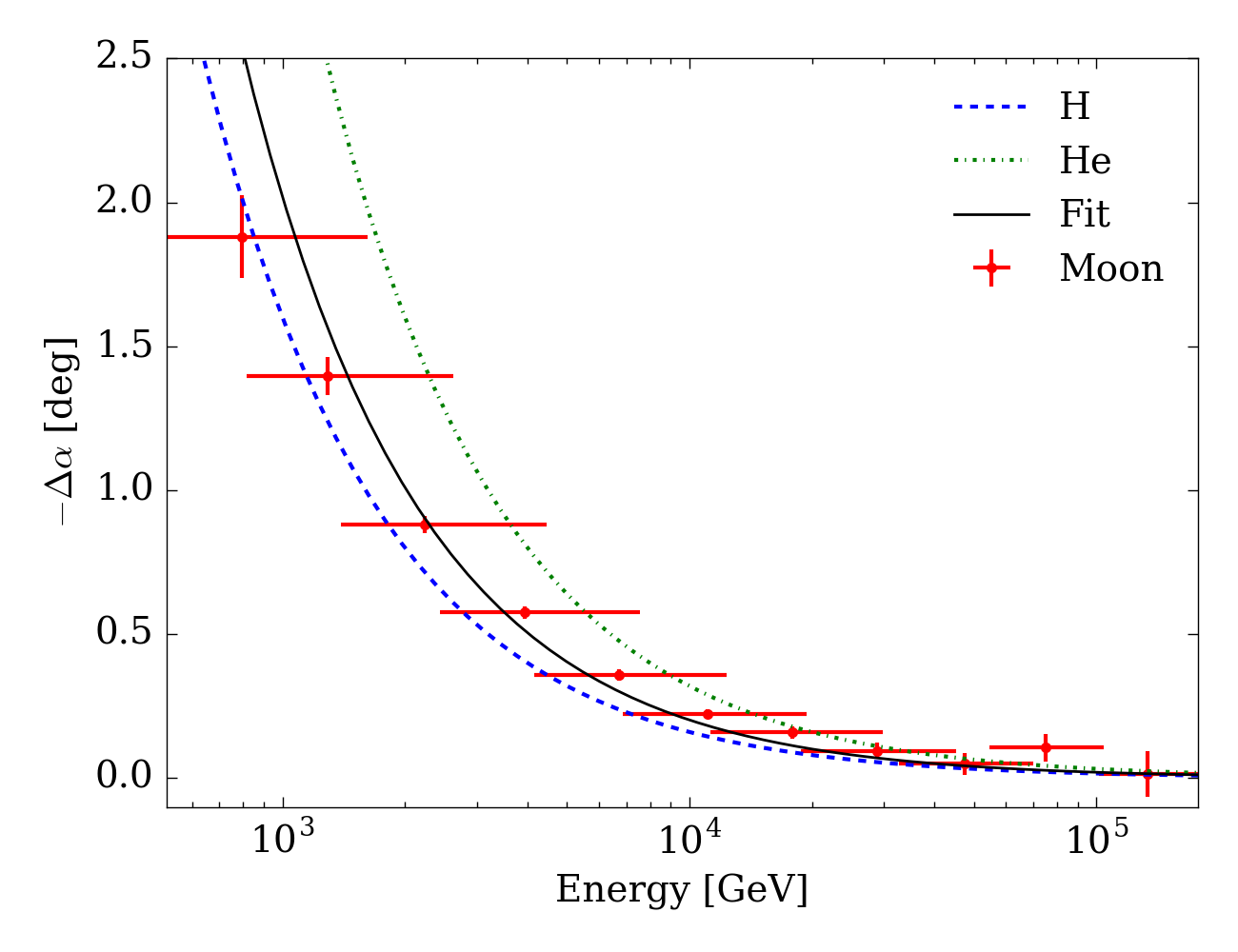}
           }

  \subfloat{
            \includegraphics[clip,width=\columnwidth]{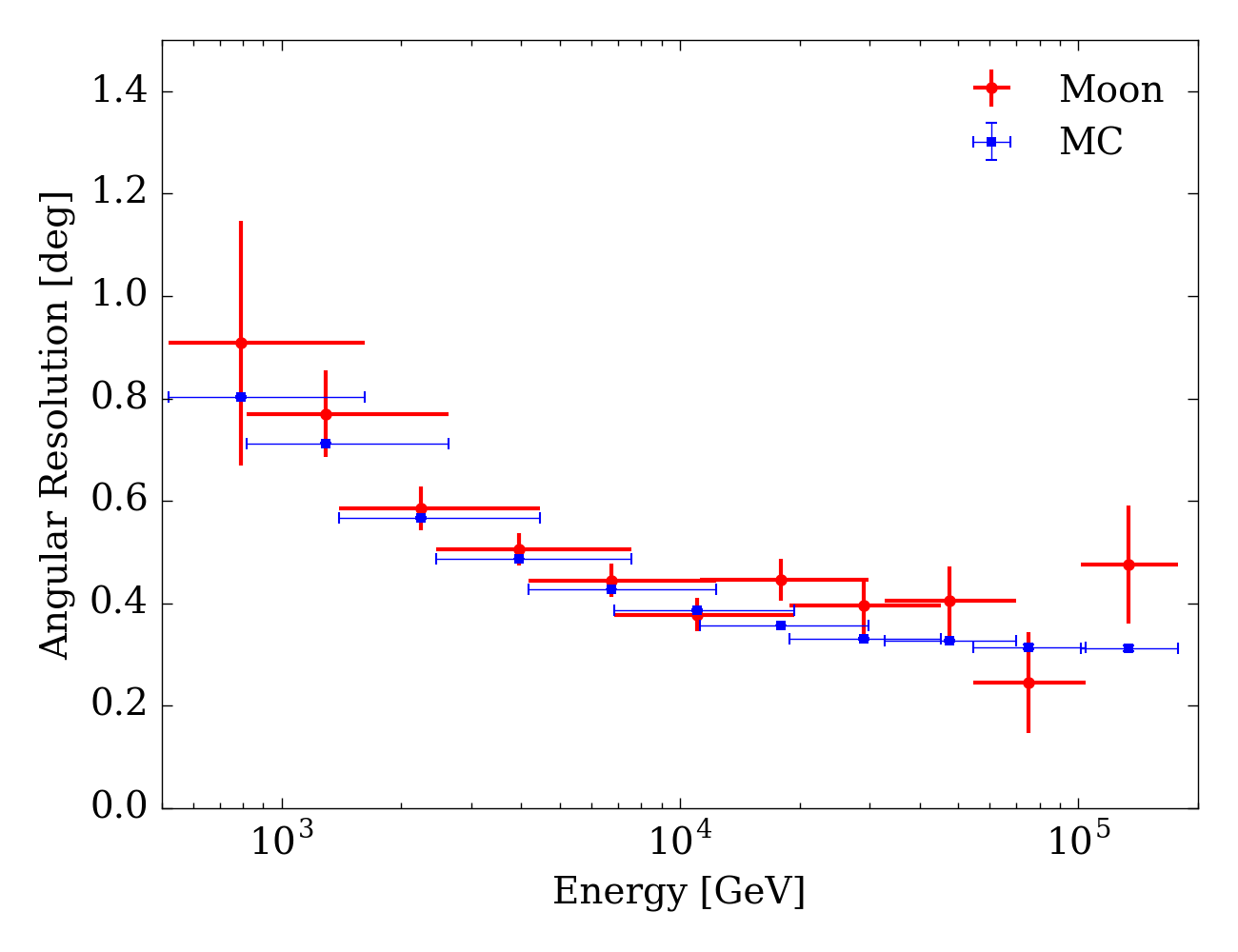}
           }
  \caption{
           Top panel: Gaussian centroid offset in right ascension $\Delta\alpha$, 
           with expectations for pure proton and pure helium hypotheses.
           Fitting the data values to equation \ref{eq:geodeflect} shown by the black curve results in an 
           estimated mean charge of $\bar{Z} = 1.25\pm0.06$. 
           Bottom panel: angular resolution from simulation compared to that estimated from using the Moon shadow width.
           The Moon points represent the fit widths 
           and are compared to the $46.6\%$ containment fraction from the simulated angular deviation distribution.
           The uncertainties on the red data points are obtained from the fit to the two-dimensional Gaussian,
           while the coordinate bins represent the estimated mean and covariance in true energy of each map.
          }

  \label{fig:moon_offset}
\end{figure}

\begin{center}
 \begin{figure*}[!htp]
  \centering
   \subfloat{
             \includegraphics[width=0.5\linewidth]{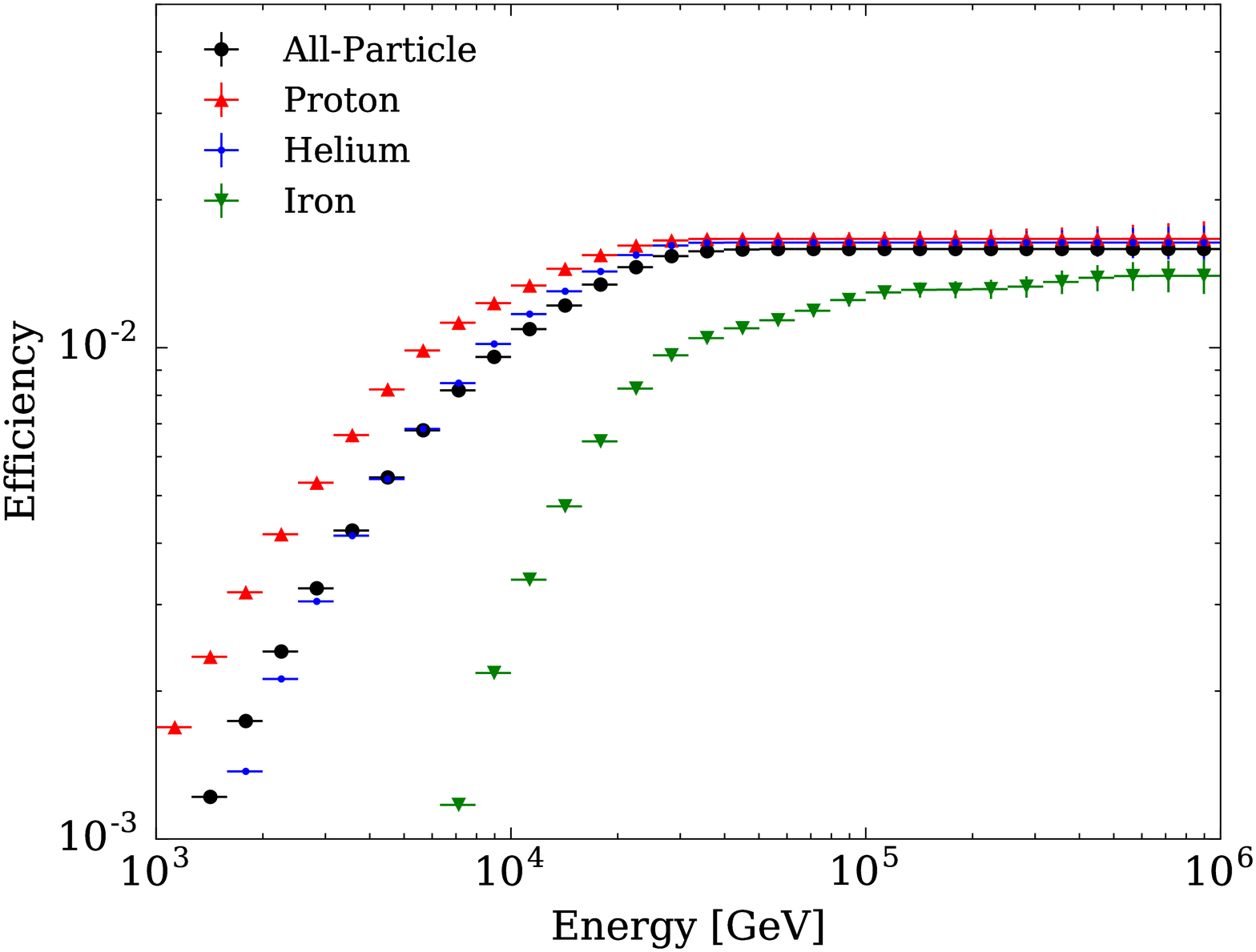}
            }
   \subfloat{
             \includegraphics[width=0.5\linewidth]{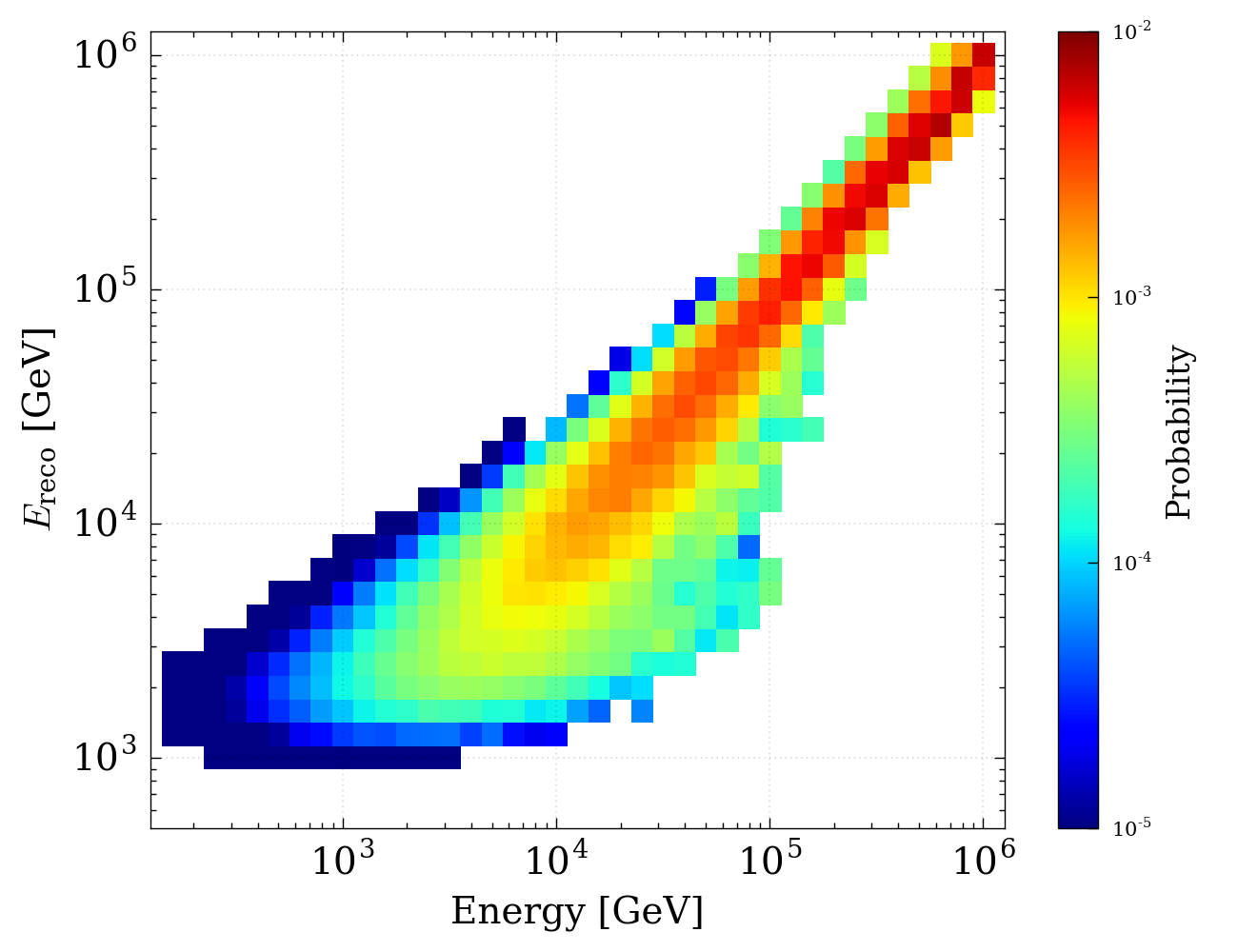}
            }
   \caption{
            The left panel shows the efficiencies $\epsilon(E)$ for all combined cosmic ray particles and 
            individually for proton, helium, and iron components.
            The energy response matrix $P(E_{\textnormal{reco}}|E)$ for all species using the composition defined 
            in table \ref{tab:crfits_table} is shown on the right.
            The deviation from the diagonal and the width of $P(E_{\textnormal{reco}}|E)$ are simply the bias and resolution, 
            respectively, already presented in figure \ref{fig:lhe-performance}.
           }
 
   \label{fig:all_part_aeff_mm}
 \end{figure*}
\end{center}

From the simulation described in section \ref{moonsim}, 
we determined the geomagnetic deflection angle $\delta\omega$ of particles with
energy $E$ and charge $Z$ arriving at HAWC to be approximately summarized by 
\begin{equation} \label{eq:geodeflect}
  \begin{split}
    \delta\omega \simeq 1.59 ^{\circ} \cdot Z \, \bigg(\frac{E}{\text{TeV}}\bigg)^{-1} \, ,
  \end{split}
\end{equation}
being inversely proportional to the particle rigidity and consistent with previous studies \cite{hawc:icrc-proc}.
We expect the deflection primarily to be in right ascension, with a small $O(0.15^{\circ} Z/E)$ deviation in declination.
Since the cosmic ray spectrum is a mixture of species and the maps are binned in energy, 
we use the composition model from section \ref{compsim} to estimate
the expected mean charge from simulation: $\bar{Z}_{\text{MC}} = 1.23\pm0.02$.
From the deviation of the observed Moon shadow, we find a mean value of $\bar{Z}=1.25\pm0.06$.
The evolution of the offset with energy is depicted in the top panel of figure \ref{fig:moon_offset}, 
along with the best fit to equation \ref{eq:geodeflect} and the expectation from simulation.
The measured width of the shadow also serves as an experimental verification of the angular resolution,
having accounted for the angular width of the Moon disc ($\sim 0.52^{\circ}$) \cite{hawc:anisotropy,hawc:dirint}.
The fraction of events contained within the $1\sigma$ region of the two-dimensional Gaussian fit to the Moon shadow is $46.6\%$,
so we identify the simulated angular deviation value having the same containment fraction.
The comparison of the angular resolution estimated from data and simulation is shown in the bottom panel of 
figure \ref{fig:moon_offset}, and we find that $\chi^2_{\text{red}} = 11.33 / 10$.

\subsection{Unfolding of the Energy Spectrum} \label{bayesunf}
Extensive air shower development is subject to inherently large fluctuations,
which result in the smearing of the primary particle's estimated energy.
Including detector effects such as limited core and angular resolutions must 
also be taken into account in order to measure the cosmic ray flux.
Following the iterative method presented in \cite{agostini}, the observed reconstructed energy 
distribution is unfolded with the estimated energy response to obtain the measured 
all-particle energy distribution.

The number of events observed in time $T$, within the solid angle $\Omega$, 
and with reconstructed energy $E_{\textnormal{reco}}$, $N(E_{\textnormal{reco}})$, is related to the true energy distribution 
$N(E)$ by the detector effective area $A_{\textnormal{eff}}(E,E_{\textnormal{reco}})$ via
\begin{equation} \label{eq:ave_acceptance}
 \begin{split}
  N(E_{\textnormal{reco}}) = 
  	\frac{1}{\Omega} \int A_{\textnormal{eff}}(E,E_{\textnormal{reco}})N(E) dE \, ,
   \end{split}
\end{equation}
where $\Omega = 2 \pi (\cos \theta_{\textnormal{min}} - \cos \theta_{\textnormal{max}})$,
and we have assumed that the detector acceptance depends only weakly on $\theta$ due to our
restricting events to be nearly vertical.
The integration is performed over the range spanned by the true energy limits of $A_{\textnormal{eff}}$.

The effective area is constructed using simulation, and can be summarized by 
\begin{equation} \label{eq:eff_area}
 \begin{split}
  A_{\textnormal{eff}}(E,E_{\textnormal{reco}}) = 
  	A_{\textnormal{thrown}} P(E_{\textnormal{reco}}|E),
   \end{split}
\end{equation}
with the simulated throwing area, 
$A_{\textnormal{thrown}} = \pi / 2 \left( \cos \theta_{\textnormal{max}}+\cos \theta_{\textnormal{min}} \right) R_{\textnormal{thrown}}^2$,
which includes a geometric factor from the zenith angle limits defined by $\theta_0$ as per Section \ref{eventselection}; 
the throwing radius $R_{\textnormal{thrown}} = 1$ km;
and $P(E_{\textnormal{reco}}|E)$ the probability for an event with energy $E$ 
to pass the event selection criteria, and reconstructed with energy $E_{\textnormal{reco}}$.

The effective area for a shower of energy $E$ is found by integrating equation \ref{eq:eff_area} over $E_{\textnormal{reco}}$,
\begin{equation} \label{eq:eff_area_E}
 \begin{split}
  A_{\textnormal{eff}}(E) = 
  	A_{\textnormal{thrown}} \epsilon (E) \, ,
   \end{split}
\end{equation}
where $\epsilon (E)$ is the efficiency to observe an event with energy $E$:
\begin{equation} \label{eq:eff_E}
 \begin{split}
	\epsilon (E) =  \int P(E_{\textnormal{reco}}|E) d E_{\textnormal{reco}} \, ,
   \end{split}
\end{equation}
and the integration limits cover the reconstructed energy range spanned by the effective area.
The object $P(E_{\textnormal{reco}}|E)$ is binned in both $E_{\textnormal{reco}}$ and $E$, forming the energy response matrix
and converting the integrals in equations \ref{eq:ave_acceptance} and \ref{eq:eff_E} to summations.
The efficiency and response matrix for all species given the event selection criteria
are shown in figure \ref{fig:all_part_aeff_mm}.
The nominal composition model from table \ref{tab:crfits_table} is also taken into account,
as species' abundances must be assumed in order to build the all-particle efficiency and response matrix,
with effects of other composition models included as systematic uncertainty of the final spectrum.

Provided a prior assumption of the energy distribution with $P(E)$,
we construct the unfolding matrix via
\begin{equation} \label{eq:bayes}
 \begin{split}
	P(E|E_{\textnormal{reco}}) =  \frac{P(E_{\textnormal{reco}}|E) P(E)}{\epsilon(E) \sum\limits_{E^{\prime}} P(E_{\textnormal{reco}}|E^{\prime})P(E^{\prime})},
   \end{split}
\end{equation}
which defines the probability of a shower with reconstructed energy $E_{\textnormal{reco}}$ 
to have been produced by a primary particle with energy $E$.
The unfolded energy distribution is given by convolving the unfolding matrix with
the reconstructed energy distribution via
\begin{equation} \label{eq:unfold}
 \begin{split}
	N(E) = \sum\limits_{E_{\textnormal{reco}}} N(E_{\textnormal{reco}}) P(E|E_{\textnormal{reco}}) \, .
   \end{split}
\end{equation}

The analysis is performed iteratively: starting with an initial prior,
the $P(E|E_{\textnormal{reco}})$ are computed via equation
\ref{eq:bayes}, providing a posterior distribution from equation
\ref{eq:unfold}.
This updated distribution is used as the subsequent prior estimate, a procedure that 
ends once variations on $N(E)$ from one iteration to the next are negligible.
For this analysis, a Kolmogorov-Smirnov test \cite{kolmogorov,smirnov} defined the convergence criterion,
where if the test statistic comparing the unfolded distributions between iterations resulted in a p-value 
less than 0.001, the unfolding was said to have converged.
The propagation of uncertainties on $N(E)$ due to the finite number of simulated and observed air-shower events 
used in the unfolding procedure follows the analytic prescription derived in Appendix B of \cite{zhampel-thesis}.
The differential flux is calculated from the final $N(E)$ according to the following relation:
\begin{equation} \label{eq:flux_derivation}
 \begin{split}
  \mathscr{F}(E) = 
  	\frac{N(E)}{T \ \Omega \ A_{\text{eff}} \ \Delta E} \, ,
   \end{split}
\end{equation}
where $T$ is the observation time, $\Omega$ is the solid angle, and $\Delta E$ 
is the energy bin width centered at $E$.

\section{Results} \label{results}

\subsection{The Unfolded Spectrum}
\begin{figure}[!htb]
 \centering
  \includegraphics[width=\linewidth]{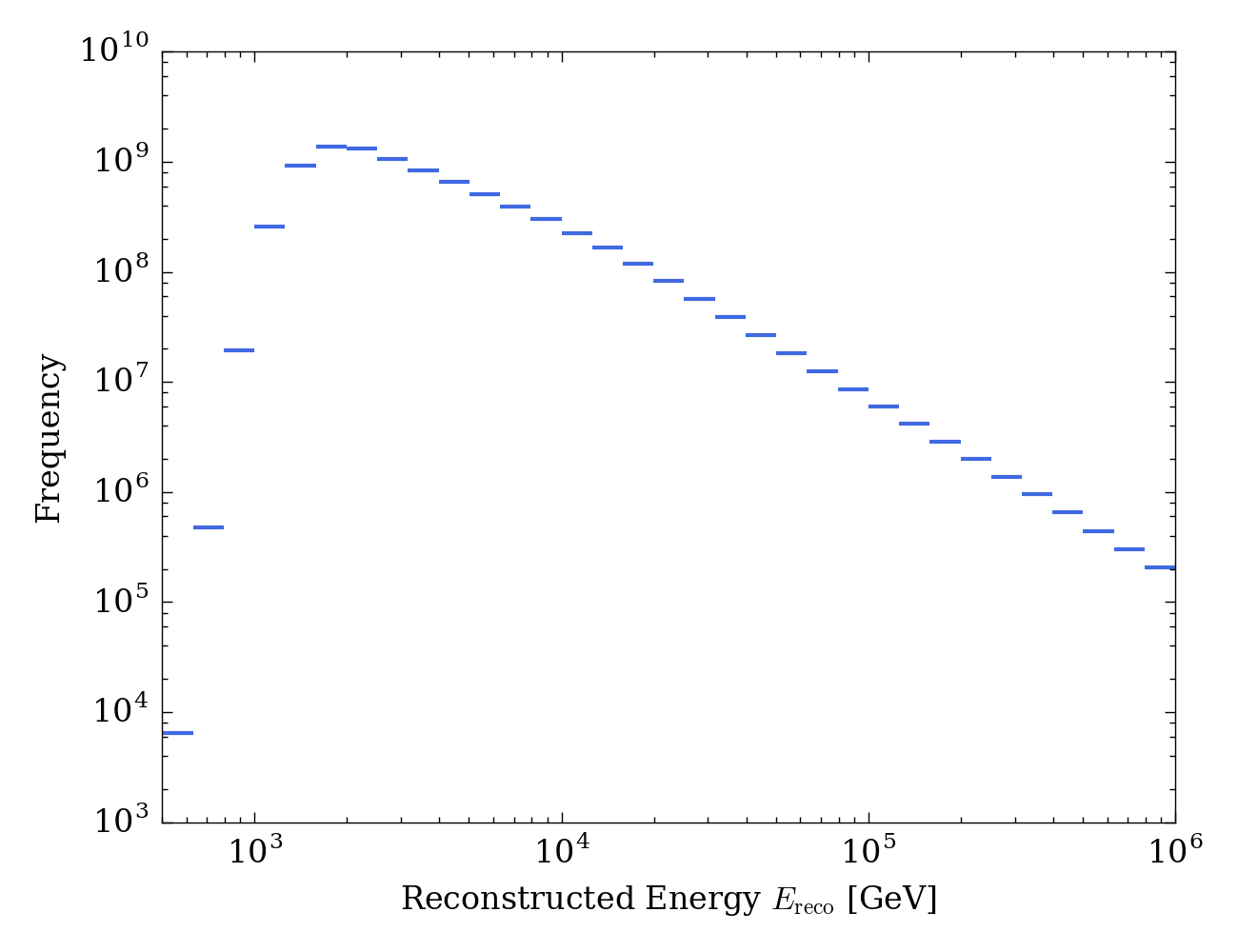}
  \caption{
           Reconstructed energy distribution $N(E_{\textnormal{reco}})$ of the 
           data sample used in this analysis.
           A subtle change in slope above $E_{\textnormal{reco}} = 2$ TeV suggests
           a change in spectral index between $10-50$ TeV in $E_{\textnormal{reco}}$.
          }
  \label{fig:reco-dist}
\end{figure}

\begin{figure*}[!htb]
 \centering
  \includegraphics[width=0.8\linewidth]{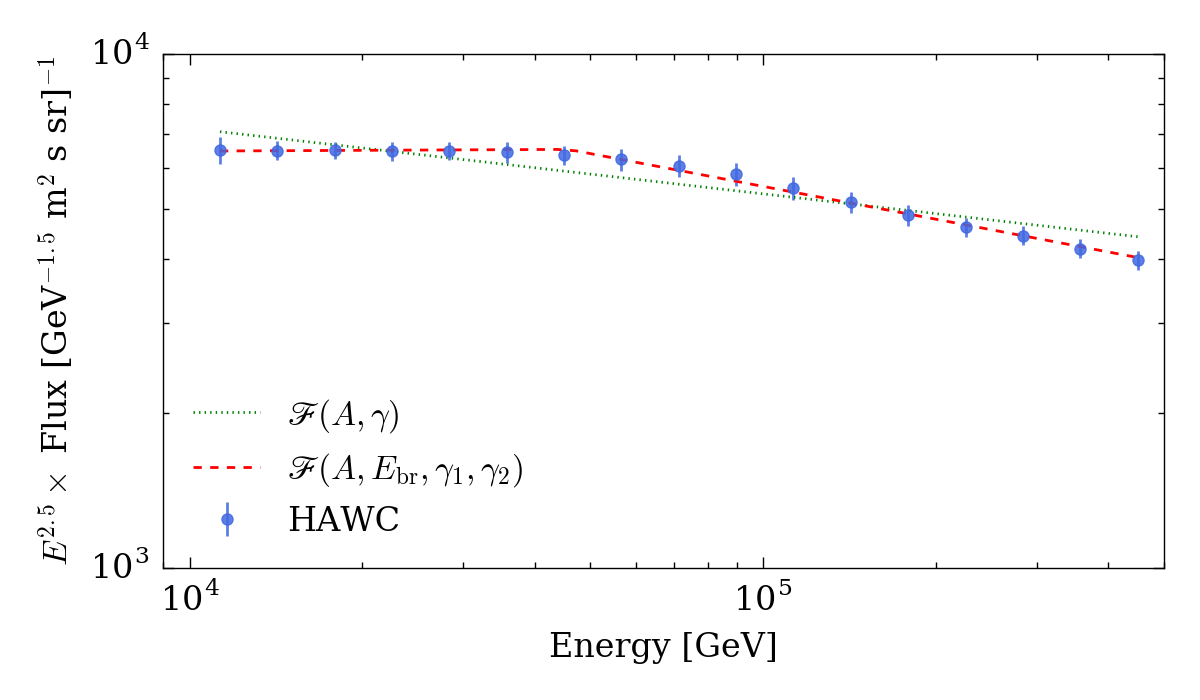}
  \caption{
            Unfolded all-particle differential energy spectrum scaled by $E^{2.5}$.
            The uncertainties visible on the data are the systematic uncertainties from the finite size of the 
            simulated data set, while the statistical uncertainties are smaller than the markersize.
            Fits to the flux using single $\mathscr{F}(A,\gamma)$ and broken $\mathscr{F}(A,E_{\text{br}}, \gamma_1,\gamma_2)$
            power law forms are also shown by the dashed lines.
            The broken power law is favored, as for $\Delta \chi^2 = 29.2$ with a difference
            of two degrees of freedom gives a p-value of $4.6\times10^{-7}$.
          }
  \label{fig:broken-pl-example}
\end{figure*}

The distribution of reconstructed energy $N(E_{\textnormal{reco}})$ provided the event 
selection criteria is shown in figure \ref{fig:reco-dist}.
Apart from the rise to the peak at $E_{\textnormal{reco}} = 2$ TeV where the 
detector efficiency is changing rapidly, there is evidence that a change in the 
spectral index is present between $10-50$ TeV in $E_{\textnormal{reco}}$.
The unfolded differential spectrum is shown in figure \ref{fig:broken-pl-example}, including only the
systematic uncertainties from the limited simulation sample used to construct the detector response function.
Due to the large data set involved, the statistical errors are not visible.
The unfolding converged in four iterations, where a non-informative prior was chosen 
as the starting distribution, and regularization via a spline fit was applied at each iteration.
Other priors of power-law form were tested with negligible effect on the final spectrum,
and similar results were found when unfolding without regularization and with power-law regularization.

The scaled differential flux reveals a feature which is not 
well described by a single power law within statistical and simulation uncertainties.
The apparent feature below $50$ TeV in true energy implies a change in the spectral index, 
so it was fit to a broken power law of the form of equation \ref{eq:fitFunc} 
and to a single power law for comparison, both shown in figure \ref{fig:broken-pl-example}.
The best fit single power law has a spectral index of $\gamma=-2.63 \pm 0.01$,
while the broken power law fit resulted in $\gamma_1 = -2.49 \pm 0.01$ and $\gamma_2 = -2.71 \pm 0.01$.
The difference in goodness of fit is $\Delta \chi^2 = 29.2$, which for a difference of two 
degrees of freedom between the two models results in a p-value of $4.6\times10^{-7}$.
Thus the broken power law is the favored model, and the fit suggests a change in spectral index of about $-0.2$ 
at a break energy of $E_{\text{br}} = (45.7 \pm 1.1)$ TeV.

Since the $N_{r40}$ variable's main effect is to select reconstructed cores, and by extension true cores, 
within the array, increasing the cut severity should improve the core resolution, 
and thus the event quality and constraints on the spectral shape.
If, for example, the under-estimated energies from poor core fits induced a feature in the spectrum, then the fit
$E_{\text{br}}$ should shift with different cuts.
Yet, we find that the fit normalizations and spectral indices for all harsher criteria agree to within 3\%, and
the location of the break energy from the fit varies by less than half the bin width ($\Delta \log E = 0.1$).
Thus we find that the spectral feature is insensitive to stronger event selection.
We also tested the response to simulated spectra and find that single power law spectra do not induce features,
whereas spectral breaks are recovered from the injection of broken power laws.

\subsection{Systematics} \label{systematics}
A thorough study of the possible systematic effects has been performed.
The main sources of systematic uncertainty considered in this work are
\begin{enumerate}[label=(\arabic*),noitemsep]

\item Effects due to the uncertainty in PMT performance properties.
      This includes the PMT charge resolution, $Q$\textsubscript{res}, and the quantum efficiency, QE.

\item Effects from the limited statistics from simulation used 
      to build the $P(E_{\text{reco}}|E)$.

\item Effects related to the assumed composition model used to build the 
      response function.

\item Effects from the hadronic interaction models used in the air shower simulations.

\end{enumerate}
Table \ref{table:sys_unc} summarizes the various contributions to the overall systematic 
uncertainty in three energy bins.
We quantified these uncertainties by building a response matrix for each of the various effects.
The observed data distribution $N(E_{\textnormal{reco}})$ was then unfolded using each model, 
and the resulting spectra were compared to the flux unfolded using the nominal response matrix 
to obtain the fractional deviation of the flux.
We do not consider uncertainty in the energy scale as a source of systematic uncertainty, though for reference 
we show in subsequent figures the shift in flux that would result from varying the energy values by $\pm10\%$.

The measurements of the all-particle energy spectrum including both systematic and statistical uncertainties
are given in table \ref{table:cr_spec}.
The total systematic uncertainty is obtained by adding the contributions from all sources in quadrature.
This is a conservative estimate following the work presented in \cite{hawc:crab}, and further study of the interplay 
of systematic effects is ongoing.
The statistical uncertainties are negligible ($\ll1\%$), due to the large ($8\times10^{9}$) event sample.

\renewcommand{\arraystretch}{1.3}
\begin{center}
 \begin{table}[!htp]
  \caption{
           Summary of systematic uncertainties.
           The contribution from each source was determined by varying that source
           independently, while holding all others fixed at their nominal values.
           The contributions from all sources are added in quadrature to conservatively
           estimate the total systematic uncertainty.
          }
  \label{table:sys_unc}
  \centering
  \begin{tabularx}{.48\textwidth}{p{2.5cm} p{1.85cm} p{1.85cm} l}
    \hline\hline
                                 & 10 TeV   & 100 TeV  & 1 PeV\\
    PMT QE                       & $\pm6\%$         &  $\pm8\%$       & $\pm9\%$\\
    PMT Q\textsubscript{res}     & $-3\%$           &  $-5\%$         & $-10\%$\\
    Simulation                   & $\pm8\%$         &  $\pm8\%$       & $\pm8\%$\\
    Composition                  & $-16/+5\%$       &  $-4/+3\%$      & $\pm3\%$\\
    Hadronic Int.                & $+5\%$           &  $+10\%$        & $-4/+2\%$\\
    \hline
    Total                        & $-20 / +  12\%$  & $-14/+15\%$     & $-20 / +13\%$ \\
    \hline\hline
  \end{tabularx}
 \end{table}
\end{center}

\begin{figure*}[!htbp]
  \includegraphics[width=0.8\textwidth]{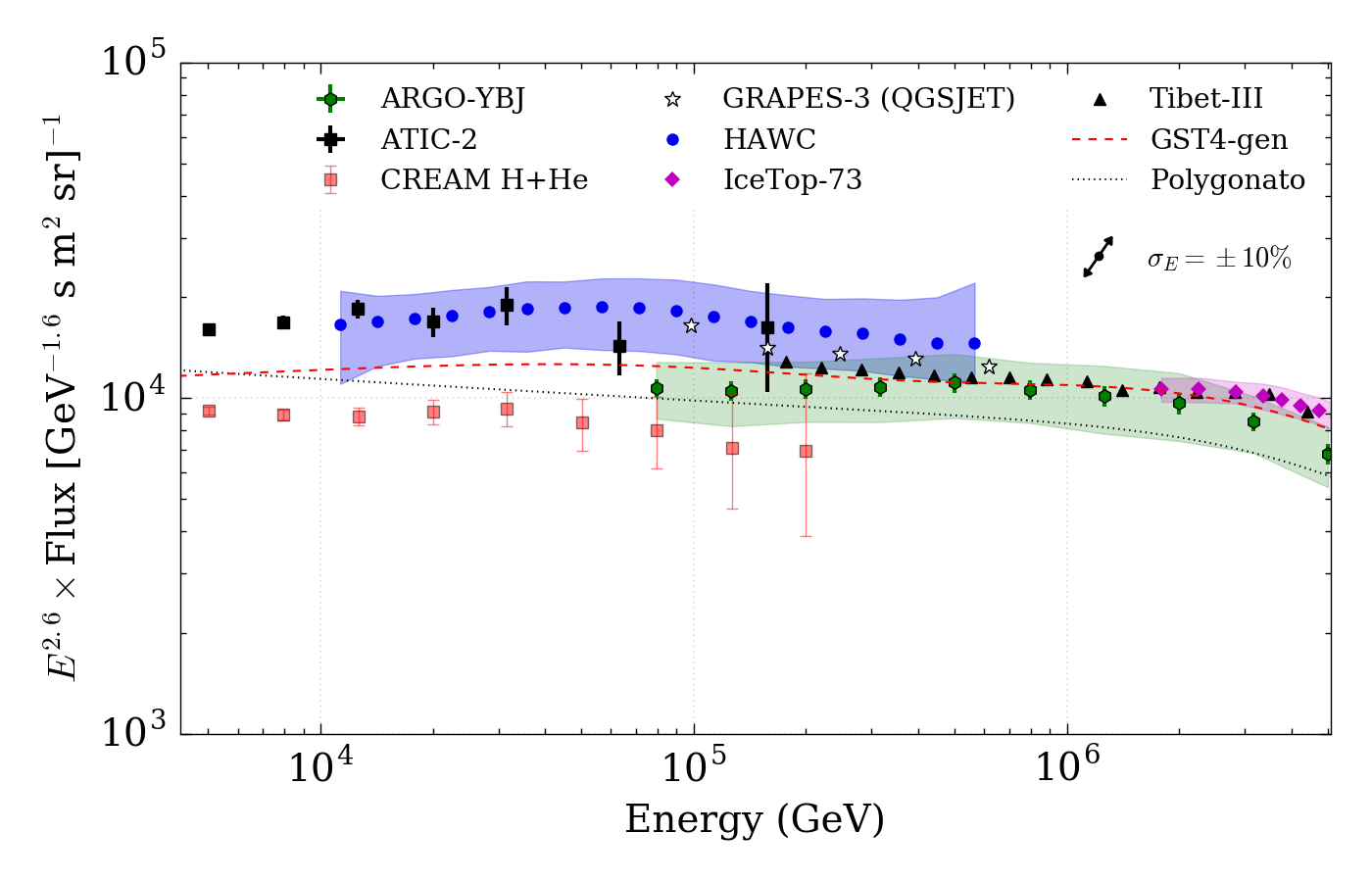}
  \caption{The differential all-particle energy spectrum measured by HAWC (blue) compared with the 
  spectra from the ARGO-YBJ \cite{argo:all-particle}, ATIC-2 \cite{atic-2:2009}, GRAPES-3 \cite{grapes-iii}, 
  IceTop \cite{icetop:allparticle}, and Tibet-III \cite{tibet:all-particle} experiments.
  The CREAM \cite{cream:2011} light component spectrum (H+He) is also included for comparison.
  The uncertainties on the ATIC-2 and CREAM measurements represent combined statistical and systematic uncertainties.
  For the HAWC, ARGO-YBJ, and IceTop spectra, the shaded regions represent the reported systematic uncertainties.
  Only ARGO-YBJ reports statistical uncertainties that are shown by visible vertical bars,
  while for the remaining air-shower array measurements, these are smaller than the respective marker size.
  The double-sided arrow indicates the shift in flux that would result from a $\pm 10 \%$ shift in the energy scale.
  The GST4-gen \cite{Gaisser2013} and Polygonato \cite{horandel} all-particle flux models are shown by the 
  red and black dashed lines, respectively.
  }
 \label{fig:crspec}
\end{figure*}

\subsubsection{PMT Charge Resolution and Quantum Efficiency}
For a fixed illumination, PMT charge measurements can vary.
We summarize this by the charge resolution, Q\textsubscript{res}, 
estimated to be between 10--25\%.
The PMTs also have an intrinsic efficiency (QE) for the combined 
conversion of an incident photon to a PE and collection of that PE, 
with typical values between 20--30\% \cite{hawc:icrc-proc}.
As a result, we vary QE and Q\textsubscript{res} values in simulation 
to obtain combined uncertainties on the energy spectra that are 
approximately 5\% below 100 TeV and no larger than 10 \% at the highest energies.

\subsubsection{Detector Simulation}
The energy resolution between 10--500 TeV given the event selection is approximately 25--50\%.
This is not considered as a systematic error, as it is taken into account in the unfolding procedure.
However, the limited statistics from simulation used in determining the $P(E_{\textnormal{reco}}|E)$ were 
taken into account as a source of uncertainty, amounting to less than $8\%$ for the all-particle spectrum.

\begin{figure*}[!htbp]
 \centering
  \includegraphics[width=0.8\textwidth]{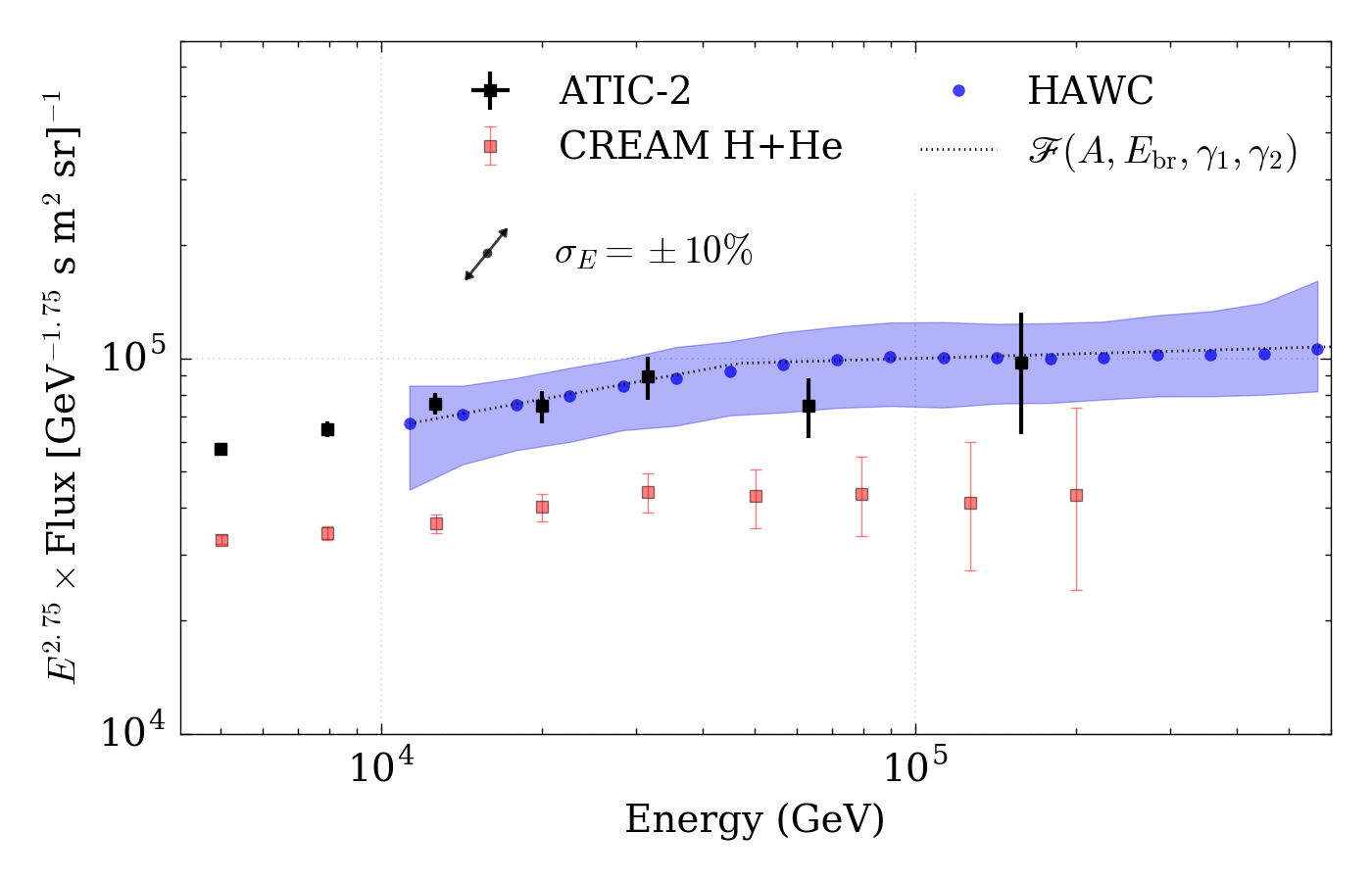}
  \caption{
           Comparison of the HAWC spectrum to the all-particle measurement by ATIC-2 \cite{atic-2:2009}
           and the light component (proton and helium) by CREAM \cite{cream:2011}.
           The energy flux is scaled by $E^{2.75}$ for ease of viewing, and
           the dashed line is 
           the best fit broken power law $\mathscr{F}(A,E_{\textnormal{br}},\gamma_1,\gamma_2)$ 
           to HAWC data from figure \ref{fig:broken-pl-example} to highlight the location of $E_{\text{br}}$.
           The double-sided arrow indicates the shift in flux that would result from a $\pm 10 \%$ shift in the energy scale.
           }
 \label{fig:crspec-direct-comp}
\end{figure*}

\subsubsection{Composition Model}

We considered two contributions to the uncertainty from the assumed composition models.
The first arises from the fit uncertainties of equation \ref{eq:fitFunc} in defining the nominal CREAM model. 
We quantified this by varying the fit parameters to within their estimated uncertainties, obtaining
the model's contribution to the uncertainty in the unfolded flux.
This amounted to less than 3\% uncertainty for all energies.
We also considered three other widely used models: H4a \cite{gaisser:hxa},
Polygonato \cite{horandel}, and the Gaisser-Stanev-Tilav model (GST4-gen) \cite{Gaisser2013}.
The H3a model also presented in \cite{gaisser:hxa} was considered; however,
the unfolded spectrum was within $<1\%$ of the unfolded spectrum using the H4a
model, so we simply quote the H4a result.
We take the full range spanned by the models in each energy bin as a conservative estimate of the 
systematic uncertainty, as we have assumed no preference for any one model.

In all, the uncertainty due to the assumed composition does not exceed $+5\%$ for all energies
and is within $-4\%$ above 100 TeV.
The greatest deviation from the nominal model comes from H4a, providing an uncertainty of $-16 \%$ at 10 TeV.
This is due to the significantly larger contribution of heavy elements ($>$He) to the model as compared to the other three.
This has the effect of reducing the efficiency (or equivalently $A_{\text{eff}}$) which can be seen in figure \ref{fig:all_part_aeff_mm}, 
since the all-particle efficiency is an abundance-weighted average for all species.
The greater presence of heavier elements also induces increased energy migration such that reconstructed
events at lower energies are promoted towards higher energies in the unfolded flux.

\subsubsection{Hadronic Interaction Model}
We also considered the systematic uncertainty from different hadronic 
interaction models by comparing the nominal simulation using QGSJet-II-03 \cite{qgsjet}
to the EPOS (LHC) \cite{epos} and SIBYLL 2.1 \cite{sibyll} high energy models.
The unfolded spectra using the EPOS model and the nominal simulation
agree to within 2\%, while the spectrum unfolded using the SIBYLL model 
is systematically higher by between 5--10 \% for all energies. 
Studies from groups such as the GRAPES-3 air shower experiment \cite{grapes-iii} found 
that for a fixed composition assumption,
the choice of hadronic interaction model influenced the relative abundance of the species 
arriving at ground level.
They found this was primarily due to model differences in determining the point of the first interaction.
As the simulated data sets for these models were smaller than the nominal set,
a more thorough analysis of the origin of these discrepancies was not possible.
Still, we include this as a source of systematic uncertainty using the observed ranges
of unfolded spectra.

\section{Discussion} \label{discussion}
A comparison of the unfolded all-particle spectrum to other recent ($<20$ yrs) experimental results 
is shown in figure \ref{fig:crspec}.
Above 100 TeV, there is agreement with the final ATIC-2 \cite{atic-2:2009} data point though it 
has statistical uncertainty greater than the HAWC systematics.
The HAWC measurement is systematically higher than measurements from other ground based 
experiments such as the GRAPES-3 \cite{grapes-iii} and Tibet-III \cite{tibet:all-particle} arrays, 
but consistent within the estimated systematic uncertainties.
Indeed, these discrepancies can be understood as arising from differences in the 
experiments' energy scale calibrations. 
For example, a 10\% systematic shift in energy results in a more dramatic $\sim30\%$ shift
in the energy scaled flux.
The ARGO-YBJ spectrum \cite{argo:all-particle} is also lower and appears harder with an index of $-2.62\pm0.03$
until $\sim700$ TeV where it softens, so the energy scaling effect does not address this discrepancy in spectral shape.
For comparison, the spectral index as measured by Tibet-III between 150--1000 TeV is $-2.68\pm0.02$,
where the uncertainties quoted are statistical and whose value is consistent with that of $\gamma_2$.

In the 10 TeV range, the HAWC spectrum is consistent with
the ATIC-2 all-particle measurement \cite{atic-2:2009}, including its rise to 30 TeV.
The slightly steeper spectrum below 50 TeV is also mirrored, though not as strongly,
in the GST4-gen model, which also depicts a downturn in the 50--60 TeV region.
The ARGO-YBJ light-component (proton and helium) measurement \cite{argo:light_2012}
does not indicate a spectral softening at these energies, having a constant slope of $-2.61 \pm 0.04$
from 5--280 TeV.
There is evidence reported by CREAM \cite{cream:2011} of a softening of
the helium spectrum between 10--30 TeV,
with both proton and helium subsequently becoming softer, though CREAM only reports
a single power law index for each species: $-2.66\pm0.02$ and $-2.58\pm0.02$, respectively.
For the nominal composition model used in this work, the broken power law fits from table 
\ref{tab:crfits_table} do not indicate spectral softening near 20--40 TeV; however, these
fits were merely used to extrapolate composition data into the $>10$ TeV regime.
The ATIC-2 combined proton and helium measurement also shows evidence of a spectral softening, 
though it peaks near 12 TeV \cite{atic-2:2009}.
Figure \ref{fig:crspec-direct-comp} shows a closer view of this region, comparing the ATIC-2,
CREAM, and HAWC spectra.
The stronger energy scaling ($E^{2.75}$) of the ordinate reveals consistent spectral forms,
with a potential kink between 20--40 TeV for ATIC-2 and a clearer break at 30 TeV for CREAM.
Since the light component comprises $\sim90\%$ of the all-particle flux, these direct detection 
experiments suggest that proton and helium are responsible for the structure observed in the HAWC measurement.

\section{Conclusions} \label{conclusion}
The special features of the HAWC experiment - full coverage, high altitude location - allow
the imaging of the front of showers induced by primaries with energies down to a few TeV, 
so far accessible only with balloon-borne experiments, and nearly up to PeV energies.
By requiring near vertical showers and applying a selection criterion based on the particle density near the fit core, 
a sample of events mainly induced by hadronic showers landing on the array has been selected.
Using a technique that relies on evaluating the signal as a function of lateral distance from the fit core, 
a proton-hypothesis based likelihood table provides the energy estimate for air-shower events.
The estimator was verified using the evolution of the cosmic ray Moon shadow, in agreement with the expected deviation 
with energy from a novel GPU-based simulation.
We built a function that connects the true energy to the estimated energy for all cosmic ray species,
assuming a CREAM-like composition model.
Using this detector response, an iterative unfolding technique has been applied to obtain the differential all-particle 
cosmic ray energy spectrum in the energy range 10--500 TeV with estimated systematics uncertainties not exceeding 20\%.

The HAWC all-particle cosmic ray spectrum exhibits agreement within estimated systematic uncertainties
with various experiments from 10--500 TeV, including evidence of a spectral break below 50 TeV.
The measurement demonstrates that HAWC can extend the reach of ground-based air-shower arrays
into the energy range covered by direct detection experiments.
Furthermore, it is with a single experimental technique that the HAWC spectrum bridges these regions.
It is also evident that HAWC has the potential to extend the spectrum up to PeV energies to probe the knee.
However, as the current event quality selection limits the range to around 500 TeV,
an improved understanding of the detector response to the highest energy events is needed.

\acknowledgments

We acknowledge the support from: the US National Science Foundation (NSF); 
the US Department of Energy Office of High-Energy Physics; 
the Laboratory Directed Research and Development (LDRD) program of Los Alamos National Laboratory; 
Consejo Nacional de Ciencia y Tecnolog\'{\i}a (CONACyT), M{\'e}xico 
(grants 271051, 232656, 260378, 179588, 239762, 254964, 271737, 258865, 243290, 132197), 
Laboratorio Nacional HAWC de rayos gamma; L'OREAL Fellowship for Women in Science 2014; 
Red HAWC, M{\'e}xico; DGAPA-UNAM (grants IG100317, IN111315, IN111716-3, IA102715, 109916, IA102917); 
VIEP-BUAP; PIFI 2012, 2013, PROFOCIE 2014, 2015; 
the University of Wisconsin Alumni Research Foundation; 
the Institute of Geophysics, Planetary Physics, and Signatures at Los Alamos National Laboratory; 
Polish Science Centre grant DEC-2014/13/B/ST9/945; 
Coordinaci{\'o}n de la Investigaci{\'o}n Cient\'{\i}fica de la Universidad Michoacana. 
Thanks to Scott Delay, Luciano D\'{\i}az and Eduardo Murrieta for technical support.

\bibliographystyle{apsrev4-1}

%

\clearpage

\onecolumngrid{

  \begin{center}
   \begin{table}
    \centering
    \caption[All-particle cosmic ray differential flux]{
             Values of the all-particle cosmic-ray energy spectrum from 10--500 TeV including uncertainties.
             The second column is the number of events unfolded, or the distribution $N(E)$.
             The label ``stat'' represents the statistical uncertainties, ``sys$_{\text{MC}}$'' is for the 
             uncertainties from the limited amount of simulation, 
             and ``sys'' represents the remaining sources of systematic uncertainty added in quadrature.
            }
    \begin{tabularx}{\textwidth}{p{3.75cm} p{5.75cm} l}
      \hline\hline
  
      $\log{E/\text{GeV}}$ & $N_{\text{events}}$ Unfolded & $\frac{dN}{dE \, d\Omega \, dt \, dA} \pm $ stat $ \pm $ sys$_{\text{MC}} + $ sys $ - $ sys [GeV s m$^2$ sr]$^{-1}$ \\
  
      \hline
 
      $4.0$ -- $4.1$ & $2.00 \times 10^{10}$ & $(4.7968 \pm 0.0002 \pm 0.5901 + 0.4288 - 0.8530)\times 10^{-7}$ \\
      $4.1$ -- $4.2$ & $1.42 \times 10^{10}$ & $(2.6922 \pm 0.0001 \pm 0.2323 + 0.2360 - 0.4467)\times 10^{-7}$ \\
      $4.2$ -- $4.3$ & $1.00 \times 10^{10}$ & $(1.5163 \pm 0.0001 \pm 0.1189 + 0.1315 - 0.2356)\times 10^{-7}$ \\
      $4.3$ -- $4.4$ & $7.08 \times 10^{9}$ &  $(8.4947 \pm 0.0007 \pm 0.7137 + 0.7352 - 1.2419)\times 10^{-8}$ \\
      $4.4$ -- $4.5$ & $5.02 \times 10^{9}$ &  $(4.7823 \pm 0.0005 \pm 0.3896 + 0.4171 - 0.6614)\times 10^{-8}$ \\
      $4.5$ -- $4.6$ & $3.54 \times 10^{9}$ &  $(2.6761 \pm 0.0003 \pm 0.2536 + 0.2377 - 0.3522)\times 10^{-8}$ \\
      $4.6$ -- $4.7$ & $2.47 \times 10^{9}$ &  $(1.4823 \pm 0.0002 \pm 0.1305 + 0.1357 - 0.1869)\times 10^{-8}$ \\
      $4.7$ -- $4.8$ & $1.71 \times 10^{9}$ &  $(8.1839 \pm 0.0015 \pm 0.8041 + 0.7830 - 0.9947)\times 10^{-9}$ \\
      $4.8$ -- $4.9$ & $1.18 \times 10^{9}$ &  $(4.4769 \pm 0.0010 \pm 0.4488 + 0.4547 - 0.5281)\times 10^{-9}$ \\
      $4.9$ -- $5.0$ & $8.03 \times 10^{8}$ &  $(2.4193 \pm 0.0007 \pm 0.2504 + 0.2655 - 0.2787)\times 10^{-9}$ \\
      $5.0$ -- $5.1$ & $5.34 \times 10^{8}$ &  $(1.2781 \pm 0.0004 \pm 0.1349 + 0.1544 - 0.1447)\times 10^{-9}$ \\
      $5.1$ -- $5.2$ & $3.56 \times 10^{8}$ &  $(6.7636 \pm 0.0027 \pm 0.6441 + 0.9164 - 0.7576)\times 10^{-10}$ \\
      $5.2$ -- $5.3$ & $2.37 \times 10^{8}$ &  $(3.5835 \pm 0.0017 \pm 0.3331 + 0.5544 - 0.3995)\times 10^{-10}$ \\
      $5.3$ -- $5.4$ & $1.59 \times 10^{8}$ &  $(1.9107 \pm 0.0011 \pm 0.1644 + 0.3430 - 0.2134)\times 10^{-10}$ \\
      $5.4$ -- $5.5$ & $1.09 \times 10^{8}$ &  $(1.0346 \pm 0.0007 \pm 0.0892 + 0.2184 - 0.1166)\times 10^{-10}$ \\
      $5.5$ -- $5.6$ & $7.25 \times 10^{7}$ &  $(5.4882 \pm 0.0047 \pm 0.4659 + 1.3920 - 0.6286)\times 10^{-11}$ \\
      $5.6$ -- $5.7$ & $4.87 \times 10^{7}$ &  $(2.9284 \pm 0.0030 \pm 0.2402 + 0.9642 - 0.3441)\times 10^{-11}$ \\

      \hline\hline
    \end{tabularx}
    \label{table:cr_spec}
   \end{table}
  \end{center}
  
}

\end{document}